\documentclass[journal]{IEEEtran}

\usepackage{graphicx}
\usepackage{latexsym}
\usepackage{amsmath}
\usepackage{amssymb}
\usepackage{epsfig}
\usepackage{cite}
\usepackage{algorithmic}
\usepackage{overpic}
\usepackage{multirow}
\usepackage[svgnames]{xcolor}
\usepackage{empheq}
\usepackage{colortbl,array}
\usepackage{multirow,bigdelim}
\usepackage{subfigure}
\usepackage{booktabs,amsmath}
\usepackage{fancyhdr, graphicx, amsmath, amssymb}
\usepackage[linesnumbered,ruled]{algorithm2e}
\usepackage{arydshln}
\usepackage[english]{babel}
\usepackage{xcolor}
\usepackage{tcolorbox}
\usepackage{soul}
\usepackage[hidelinks]{hyperref}
\hypersetup{breaklinks=true}
\newtheorem{lem}{Lemma}
\newtheorem{thm}{Theorem}

\newtheorem{asm}{Assumption}
\newtheorem{definition}{Definition}
\newtheorem{rem}{Remark}
\newtheorem{prop}{Proposition}
\newtheorem{ex}{Example}
\colorlet{shadecolor}{FloralWhite!60}
\colorlet{titleshade}{OliveDrab!15}
\newsavebox{\mysaveboxM}
\newsavebox{\mysaveboxT}
\newcommand*\Garybox[2][Example]{
\sbox{\mysaveboxM}{#2}
\sbox{\mysaveboxT}{\fcolorbox{black}{titleshade}{\enspace#1\enspace}}%
\sbox{\mysaveboxM}{
\parbox[b][\ht\mysaveboxM+.5\ht\mysaveboxT+.5\dp\mysaveboxT][b]{%
\wd\mysaveboxM}{#2}
}%
\sbox{\mysaveboxM}{
\fcolorbox{black}{shadecolor}{%
\makebox[\linewidth-4.05em]{\usebox{\mysaveboxM}}%
}
}
\usebox{\mysaveboxM}%
\makebox[0pt][r]{%
\makebox[\wd\mysaveboxM][c]{%
\raisebox{\ht\mysaveboxM-0.5\ht\mysaveboxT
+0.5\dp\mysaveboxT-0.5\fboxrule}{\usebox{\mysaveboxT}}%
}
}
}

\begin{document}
\title{Novel Stealthy Attack and Defense Strategies for Networked Control Systems}
\author{Yanbing~Mao, Hamidreza~Jafarnejadsani, Pan~Zhao, Emrah~Akyol,
        and Naira~Hovakimyan
\thanks{Y.~Mao, P.~Zhao and N.~Hovakimyan are with the Department of Mechanical Science and Engineering, University of Illinois at Urbana--Champaign, Urbana, IL, 61801 USA (e-mail: \{ybmao, panzhao2, nhovakim\}@illinois.edu).}
\thanks{H.~Jafarnejadsani is with the Department of Mechanical Engineering, Stevens Institute of Technology, Hoboken, NJ,
07310 USA (e-mail: hjafarne@stevens.edu).}
\thanks{E.~Akyol is with the Department of Electrical and Computer Engineering, Binghamton University--SUNY, Binghamton, NY,
13902 USA (e-mail: eakyol@binghamton.edu).}
\thanks{Parts of this paper were presented at the 58th IEEE Conference on Decision and Control, 2019 \cite{mhp2019}. This work is supported in part by NSF (award numbers CMMI-1663460 and ECCS-1739732), and Binghamton University--SUNY, Center for Collective Dynamics of Complex Systems ORC grant.}}

\maketitle
\begin{abstract}
This paper studies novel attack and defense strategies, based on a class of stealthy attacks, namely the zero-dynamics attack (ZDA), for multi-agent control systems. ZDA  poses a formidable security challenge since its attack signal is hidden in the null-space of the state-space representation of the control system and hence it can evade conventional detection methods. An intuitive defense strategy builds on changing the aforementioned representation via switching through a set of carefully crafted topologies. In this paper, we propose realistic ZDA variations where the attacker is aware of this topology-switching strategy, and hence employs the following policies to avoid detection: (i) pause, update and resume ZDA according to the knowledge of switching topologies;  (ii) cooperate with a concurrent stealthy topology attack that alters network topology at switching times, such that the original ZDA is feasible under the corrupted topology. We first systematically study the proposed ZDA variations, and then develop defense strategies against them under the realistic assumption that the defender has no knowledge of attack starting, pausing, and resuming times and the number of misbehaving agents. Particularly, we characterize conditions for detectability of the proposed ZDA variations, in terms of the network topologies to be maintained, the set of agents to be monitored, and the measurements of the monitored agents that should be extracted, while simultaneously preserving the privacy of the states of the non-monitored agents. We then propose an attack detection algorithm based on the Luenberger observer, using the characterized detectability conditions. We provide numerical simulation results to demonstrate our theoretical findings.
\end{abstract}

\IEEEpeerreviewmaketitle
\section{Introduction}
\IEEEPARstart{C}{oordination} and control of networked systems is a well-studied theoretical problem (see e.g., \cite{jadbabaie2003coordination,fax2004information}) with many practical applications including distributed optimization~\cite{nedic2009distributed}, power sharing for droop-controlled inverters in islanded microgrids~\cite{lu2015consensus}, clock synchronization for sensor networks~\cite{li2006global},  as well as connected vehicles~\cite{ren2007distributed}, spacecrafts~\cite{abdessameud2009attitude}, and
electrical power networks~\cite{johnson2014synchronization}. 

Security concerns regarding the aforementioned networked systems pose a formidable threat to their wide deployment, as highlighted by the recent incidents including distributed denial-of-service (DDOS) attack on Estonian web sites~\cite{nazario2009politically} and Maroochy water breach \cite{slay2007lessons}.  The ``networked" aspect  exacerbates the difficulty of securing these systems, since centralized measurement (sensing) and control are not feasible for such large-scale systems~\cite{F13}, and hence require the development of decentralized approaches, which are inherently prone to attacks.  Particularly, a special class of stealthy attacks, namely the ``zero-dynamics attack" (ZDA), poses a significant security challenge~\cite{MP15,hl18,MP18}. The main idea behind ZDA is to hide the attack signal  in the null-space of the state-space representation of the control system so that it cannot be detected by applying conventional detection methods on the observation signal.  The objective of such an attack can vary from manipulating the controller to accept false data that would yield the system towards a desired (e.g., unstable) state to maliciously altering system dynamics (topology attack) to affect the system trajectory.

Recent research efforts have focused on variations of ZDA for systems with distinct properties. For stochastic cyber-physical systems, Park et al.~\cite{park2016} designed a robust ZDA, where the attack-detection signal is guaranteed to stay below a threshold over a finite horizon. In~\cite{kim2016zero}, Kim et al. proposed a discretized ZDA for the sampled-data control systems, where the attack-detection signal is constant zero at the sampling times. Another interesting line of research pertains to developing defense strategies \cite{n11,n12,F13,weerakkody2017robust,chen2017protecting}.  For example, Jafarnejadsani et al.~\cite{hl18} proposed a multi-rate $\mathcal{L}_{1}$ adaptive controller that can detect ZDA in sampled-data control systems by removing certain unstable zeros of discrete-time systems. Back et al.~\cite{back2017enhancement}  used generalized hold strategy to mitigate the impact of ZDA.

Most of the prior work on defense strategies for the original ZDA in networked systems builds on rather restrictive assumptions regarding the connectivity of network topology and the number of the misbehaving agents (i.e., the agents under attack) \cite{n11,n12,F13,weerakkody2017robust}. Teixeira et al.~\cite{T12} showed that the strategic changes in system dynamics could be used by defender to detect ZDA. But the defense strategy requires
the attack-starting times to be the initial time and known to defender, and the attacker has no capability of inferring the changed system dynamics. In other words, the defense strategy fails to work if the stealthy attack strategy is based on the newly inferred system dynamics. As a first step towards a practical ZDA defense strategy,  in \cite{ma1811},  strategic topology switching is proposed. This strategy is  motivated by the feasibility  of controlling communication topology driven due to recent developments in mobile computing, wireless communication and sensing~\cite{hartenstein2008tutorial,mazumder2011wireless}. We note, in passing, that the idea of using the changes in the state-space dynamics to detect ZDA first appeared in \cite{T12}, albeit a realistic mechanism (e.g., switching the system topology) to achieve that objective was only very recently studied in \cite{ma1811}. However, the  defense strategy in  \cite{ma1811}  still relies on a naive attacker that does not take the topology switching strategy of the defender into account.

In this paper, we systematically address this practically important problem:  what kind of ZDA strategies can an informed attacker design against a topology-switching system and what are the optimal defense strategies, beyond switching the topology, against such intelligent attacks? We note that we study these questions under realistic assumptions on the capabilities of the defender, i.e., we assume that the defender does not know the start, pause and resume times of the attack or the number of misbehaving agents.  We also assume that  the attacker is aware of the strategic changes in system dynamics. Moreover, we assume that the defender has to preserve the privacy of the outputs of the non-monitored agents, since it is assumed that the attacker has access to the sensor outputs. The following example from coordination control illustrates our motivation to impose this privacy constraint.

For the coordination control of multi-agent systems, e.g., the connected autonomous vehicles, the data of initial positions and velocities can be used by the adversary to estimate target location~\cite{xue2014security}, and the individual initial positions include individual home-base locations. Once the attacker has access to the outputs of monitored agents and the system is observable, the attacker can use current available data to infer the global initial condition and global real-time system state. From a perspective of stealthy topology attack design (e.g., topology attack in smart grids~\cite{kim2013topology} and software-defined networks~\cite{skowyra2018effective}), the attacker needs (estimated) real-time data of some agents' state to decide the target connection links to attack. Unfortunately, the inferred global real-time system state implies the largest scope of attackable connection links is exposed to the attacker. To reduce the feasible area of target links for ZDA in cooperation with a stealthy topology attack, monitored outputs have to be constrained to be unobservable to  preserve the privacy of non-monitored agents' real-time states, consequently, the global system state and global initial condition.

Throughout this paper, we focus on the following policies which can be used by the attacker to evade detection:
\begin{enumerate}
  \item intermittently pause attack if the incoming topology is unknown, and update (if necessary) and resume attack after the newly activated topology is inferred (intermittent ZDA).
  \item cooperatively work with a stealthy topology attack, such that the original ZDA policy continues to be feasible under the corrupted topology (cooperative ZDA).
\end{enumerate}

In this paper, we develop integrated defense strategies for both intermittent and cooperative ZDAs, in the presence of privacy considerations. More specifically, we develop defense strategies to address the following questions: what network topology should be maintained, which agents should be monitored and what measurements the monitored agents should output, such that the {\it intermittent} and {\it cooperative} ZDA variants are detectable, and at the same time, the privacy of non-monitored agents' real-time states are preserved? Based on the answers of the questions above, we next propose a strategic topology-switching algorithm to detect the ZDA.

The contributions are summarized as follows.
\begin{itemize}
 \item To evade conventional detection methods that rely on naive attacker, we propose two ZDA variations: intermittent and cooperative ZDAs, where the attacker is aware of the defense strategy and has practical capability of inferring switching topologies.
 \item We systematically study the policies of ZDA variations that the attacker follows to devise stealthy attacks that lay the foundation for the novel defense strategies.
  \item We characterize conditions for detectability of the proposed ZDA variations, in terms of the network topologies to be maintained, the set of agents to be monitored, and the measurements of the monitored agents that should be extracted.
  \item Under the privacy-preserving constraint of non-monitored agents' states,  we propose a strategic topology-switching algorithm for attack detection that is based on the detectability of ZDA variations using the Luenberger observer. The advantages of this approach include:
      \begin{itemize}
        \item in achieving consensus and tracking real systems in the absence of attacks, it has no constraint on the magnitudes of coupling weights and observer gains;
        \item in detecting ZDA variations, it allows the defender to be unaware of attack starting, pausing, and resuming times and the number of misbehaving agents;
        \item in detecting ZDA variations, only one monitored agent is sufficient for intermittent ZDA and only two monitored agents are sufficient for cooperative ZDA.
      \end{itemize}
\end{itemize}

This paper is organized as follows. We present the preliminaries and the problem formulation in Sections II and III, respectively. In Section IV, we analyze the proposed ZDA variations. In Section V, we characterize the conditions for detectability of these ZDA variations.  Based on this characterization, we develop an attack detection algorithm in Section VI. Numerical simulation results are provided in Section VII, and the concluding remarks and the future research directions are discussed in Section VIII.

\section{Preliminaries}
\subsection{Notation}
We let $\mathbb{R}^{n}$ and $\mathbb{R}^{m \times n}$ denote the set of $\emph{n}$-dimensional real vectors and the set of $m \times n$-dimensional real matrices, respectively. Let $\mathbb{C}$ denote the set of complex numbers. $\mathbb{N}$ represents the set of the natural numbers and $\mathbb{N}_{0}$ = $\mathbb{N}$ $\cup$ $\left\{ 0 \right\}$. Let $\mathbf{1}_{n \times n}$ and $\mathbf{0}_{n \times n}$ be the $n \times n$-dimensional identity matrix and zero matrix, respectively. $\mathbf{1}_{n} \in \mathbb{R}^{n}$ and $\mathbf{0}_{n} \in \mathbb{R}^{n}$ denote the vector with all ones and the vector with all zeros, respectively. The superscript `$\top$' stands for matrix transpose. ${\mu _P}\left( A \right)$ denotes the induced $P$-norm matrix measure of $A \in {\mathbb{R}^{n \times n}}$, with $P > 0$, i.e., ${\mu _P}\left( A \right) = \frac{1}{2}\mathop {\max }\limits_{i = 1, \ldots ,n} \left\{ {{\lambda _i}\left( {{P^{1/2}}A{P^{ - 1/2}} + {P^{ - 1/2}}{A^\top}{P^{1/2}}} \right)} \right\}$. $\ker \left( Q \right) \triangleq \left\{ {y: Qy = {\mathbf{0}_n}}, Q \in \mathbb{R}^{n \times n} \right\}$, $A^{-1}\mathbb{F} \triangleq$  $\left\{ {y: Ay \in \mathbb{F}}\right\}$. Also, $\left| \cdot \right|$ denotes the cardinality of a set, or the modulus of a number. ${\mathbb{V}} \backslash \mathbb{K}$ describes the complement set of $\mathbb{K}$ with respect to $\mathbb{V}$. $\lambda_{i}\left( M \right)$ is $i^{\emph{\emph{th}}}$ eigenvalue of matrix $M$. $x^{(b)}(t)$ stands for the $b^{\emph{\emph{th}}}$-order time derivative of $x(t)$. For a matrix $W \in \mathbb{R}^{n \times n}$, $W^{k}$, ${\left[ {{W}} \right]_{i,j}}$, ${\left[ {{W}} \right]_{i,:}}$, and ${\left[ {{W}} \right]_{a:b,c:d}}$ denote the $k^{\emph{\emph{th}}}$ power of $W$, the element in row $i$ and column $j$,  the $b^{\emph{\emph{th}}}$ row, and the sub-matrix formed by the entries in the $a^{\emph{\emph{th}}}$ through $b^{\emph{\emph{th}}}$ row and the $c^{\emph{\emph{th}}}$ through $d^{\emph{\emph{th}}}$ column of $W$, respectively.

The interaction among $n$ agents is modeled by an undirected graph $\mathrm{G} \triangleq (\mathbb{V}, \mathbb{E})$, where
$\mathbb{V}$ $\triangleq$ $\left\{ {1,2, \ldots, n} \right\}$ is the set of vertices that represents $n$ agents and $\mathbb{E} \subseteq \mathbb{V} \times \mathbb{V}$ is the set of edges of the graph $\mathrm{G}$. The weighted adjacency matrix $\mathcal{A} = \left[ {{a_{ij}}} \right]$ $\in \mathbb{R}^{n \times n}$ of the graph $\mathrm{G}$ is defined as $a_{ij} = a_{ji} > 0$ if $(i, j) \in \mathbb{E}$, and $a_{ij} = a_{ji} = 0$ otherwise. Assume that there are no self-loops, i.e., for any ${i} \in \mathbb{V}$, $a_{ii} = 0$.  The Laplacian matrix of graph $\mathrm{G}$ is defined as $\mathcal{L} \triangleq \left[ {{l_{ij}}} \right] \in {\mathbb{R}^{n \times n}}$, where ${l_{ii}} \triangleq \sum\limits_{j = 1}^n {{a_{ij}}}$, and ${l_{ij}} \triangleq - {a_{ij}}$ for $i \neq j$. The diameter $m$ of a graph is the longest shortest unweighted path between any two vertices in the graph.

\vspace{-0.20cm}
\subsection{Definitions}
\vspace{-0.1cm}
A second-order system consists of a population of $n$ agents whose dynamics are governed by the
following equations:
\vspace{-0.1cm}
\begin{subequations}\label{eq:MA_sys}
\begin{align}
{\dot x_i}\left( t \right) &= {v_i}\left( t \right),\label{eq:oox1}\\
{{\dot v}_i}\left( t \right) &=  u_{i}(t),\hspace{0.5cm}i = 1, \ldots ,n\label{eq:oox2}
\end{align}\label{eq:ooiio}\end{subequations}
where $x_{i}(t) \in \mathbb{R}$ is the position, $v_{i}(t) \in \mathbb{R}$ is the velocity, and $u_{i}(t) \in \mathbb{R}$ is the local control input.  The broad applications of its coordination control is the main motivation of this paper considering the model~(\ref{eq:ooiio}), see e.g., \cite{tanner2007flocking,rahili2017distributed,olfati2006flocking,lawton2003decentralized}. For coordination control, we consider the more representative average consensus.

We recall the definitions of consensus and ZDA to review the control objective and the attack policy.
\begin{definition}
\cite{mei2016distributed} The agents in the system \eqref{eq:ooiio} are said to achieve the asymptotic consensus with final zero common velocity if for any initial condition:
\begin{align}
\mathop {\lim }\limits_{t \to \infty } | {{x_i}\left( t \right) \!-\! {x_j}\left( t \right)} | \!=\! 0 \hspace{0.1cm}\emph{\emph{and}} \mathop {\lim }\limits_{t \to \infty } | {{v_i}\left( t \right)} | \!=\! 0, \forall i, j \in \mathbb{V}. \label{eq:defc}
\end{align}\label{okokerer}
\end{definition}
\begin{definition}~\cite{teixeira2012attack,F13} Consider the system (with proper dimension) in the presence of attack signal $\breve{g}(t)$:
\begin{subequations}
\begin{align}
\dot{\breve{z}}\left( t \right) &= A\breve{z}\left( t \right) + B\breve{g}(t),\\
\breve{y}\left( t \right) &= C\breve{z}\left( t \right) + D\breve{g}(t).
\end{align}\label{eq:sl2}\end{subequations}
The attack signal ${\breve{g}}(t) = ge^{\eta t}$ is a \emph{zero-dynamics attack} if there exist a scalar $\eta \in \mathbb{C}$, and nonzero vectors $\mathbf{z}_{0}$ and ${g}$, that satisfy
\begin{align}
\left[\!
    \begin{array}{c}
        \mathbf{z}_{0}\\ \hdashline[2pt/2pt]
        { - g}
    \end{array}
\!\right] &\in \ker \left(\left[\!
    \begin{array}{c;{2pt/2pt}c}
        {\eta \mathbf{1}_{{n} \times {n}} - A} \!&\! B\\ \hdashline[2pt/2pt]
        -C \!&\! D
    \end{array}
\!\right]\right).\label{eq:czeo}
\end{align}
Moreover, the states and observed outputs of system \eqref{eq:sl1} satisfy
\begin{align}
 \breve{y}\left( t \right) &= y\left( t \right), t \ge 0 \label{eq:rs1}\\
\breve{z}\left( t \right) &= z\left( t \right) + \mathbf{z}_{0}{e^{\eta  {t } }},\label{eq:rs2}
\end{align}
where $y\left( t \right)$ and $z\left( t \right)$ are the output and state of the system \eqref{eq:sl2} in the absence of attacks, i.e., the dynamics:
\begin{subequations}
\begin{align}
\dot{{z}}\left( t \right) &= A{z}\left( t \right),\\
{y}\left( t \right) &= C{z}\left( t \right) .
\end{align}\label{eq:sl1}\end{subequations}
\label{thm:zods}
\end{definition}

\vspace{-0.9cm}
\subsection{Control Protocol}
We borrow a control protocol that involves topology switching from~\cite{xie2006consensus,mei2016distributed} to achieve the consensus~(\ref{eq:defc}) for the agents in system \eqref{eq:MA_sys}:
 \vspace{-0.1cm}
\begin{align}
u_i (t)  =  - v_i (t) + \sum\limits_{j \in \mathbb{V}} {a_{ij}^{\sigma \left( t \right)}\left( {{x_j}\left( t \right) - {x_i}\left( t \right)} \right)}, i \in \mathbb{V} \label{eq:lci}
\end{align}
where $\sigma (t):[t_{0},\infty ) \to \mathbb{S} \triangleq \{1, \ldots, \mathrm{s}\}$, is the switching signal of the interaction topology of the communication network; $a^{\sigma(t)}_{ij}$ is the entry of the weighted adjacency matrix that describes the activated topology of communication graph.

\section{Problem Formulation}
We let $\mathbb{K}$ $\subseteq$ $\mathbb{V}$ denote the set of misbehaving agents, i.e., the agents whose local control inputs are under attack. For simplicity, we let the increasingly ordered set $\mathbb{M} \triangleq \left\{ {1,2, \ldots } \right\}$ $\subseteq \mathbb{V}$ denote the set of monitored agents for attack detection.

We make the following assumptions on the attacker and defender throughout this paper.
\begin{asm}
The attacker
\begin{enumerate}
  \item is aware that the changes in system dynamics are used by the defender (system operator);
  \item knows the initial topology, output matrix and switching times;
  \item needs a \emph{non-negligible} time to exactly infer the newly activated topology, compute and update attack strategy;
  \item records the newly inferred topology into  memory;
  \item knows the outputs of monitored agents in $\mathbb{M}$.
\end{enumerate}
\label{thm:att}
\end{asm}
\begin{asm}
 The defender
 \begin{enumerate}
   \item designs the switching times and switching topologies;
  \item chooses candidate agents to monitor, i.e., the monitored agent set $\mathbb{M}$, for attack detection;
  \item  has no knowledge of the attack starting, pausing and resuming times, and the misbehaving agents.
 \end{enumerate}\label{thm:auatt}
\end{asm}
\vspace{-0.00cm}
\begin{rem}
In Assumption 1, the assumed attacker's capability 2) is motivated by recent incidents, see e.g., the revenge sewage attack (cyber attack) that led to the Maroochy water breach, where the attacker had previously installed the industrial control systems for the water service network (consequently, he knew the control protocol and the locations of sensors)~\cite{software}.
\end{rem}
\begin{rem}
Strategically changing the system dynamics has been demonstrated to be an effective approach to detect system-based stealthy attacks, see e.g., ZDA~\cite{T12,ma1811} and $\mathcal{C}_{k}/\mathcal{C}$ stealthy attacks~\cite{teixeira2014security}. The core idea behind this defense strategy is the intentional generation of mismatch between the models of the attacker and the defender. Specifically, the attacker uses the original system dynamics to make the stealthy attack decision (i.e., the computation \eqref{eq:czeo}) before the system starts to operate, while the defender strategically
changes the system dynamics at some operating point in time. However, from the attacker's perspective, it is practical to become aware of this defense strategy, and hence try to infer the changed system dynamics to update  the stealthy attack strategy and evade detection. This motivates the awareness capability 1) in Assumption \ref{thm:att}.
\end{rem}
\begin{rem}
Although the switching topologies are kept confidential from  attackers, the developed topology inference algorithms \cite{van2019topology, all19} enable the attacker to exactly infer the switching topologies from observation signals. Even with the global ability of observing all agents' states, the inference algorithms need to collect the state data over a time interval to obtain an exact topology solution, which explains the imposed \emph{non-negligible} time in capability 3) in Assumption \ref{thm:att}.
\end{rem}
\begin{rem}
Since the sensor devices are embedded within an environment, they are frequently vulnerable to local eavesdropping, which is the motivation of capability 5) in Assumption \ref{thm:att}.
The ZDA policy \eqref{eq:czeo} shows that the attacker does not need the capability 5) to obtain a feasible attack strategy consisting of the false data $\mathbf{z}_{0}$, and the parameters $g$ and $\eta$ of attack signal ${\breve{g}}(t)$. However, when ZDA seeks cooperation with a stealthy topology attack in response to strategic topology switching defense, then the attacker needs the real-time outputs indicated by the capability to identify the target links to attack.
\end{rem}
\begin{rem}
As analyzed in \cite{mhp2019}, the defense strategy of strategically changing system dynamics~\cite{T12} implicitly assumes that the attack-starting time must be the initial time and known to the defender. The capability 3) of defender in Assumption \ref{thm:auatt} removes this unrealistic assumption.
\end{rem}

\vspace{-0.00cm}
\subsection{Topology Switching Strategy} \label{III.B}
The building block of our defense strategy is periodic topology switching, i.e., there exists a period $\tau$ such that
\vspace{-0.00cm}
\begin{align}
\sigma \left( t \right) = \sigma \left( {t + \tau} \right) \in \mathbb{S}. \label{eq:psd}
\end{align}
\vspace{-0.70cm}
\begin{itemize}
  \item We note that \eqref{eq:psd} implies the building block belongs to the time-dependent topology switching. The critical reason that we do not consider state-dependent switching is the attack signals injected into control input may generate a Zeno behavior \cite{1470118} that renders the control protocol \eqref{eq:lci} infeasible.
  \item If the topology switching is random, the defender needs to often send the generated ``random" information of network topology to the detector/estimator/observer in the cyber layer as well, which will be subject to a cyber topology attack (incorrect information of network topology is transmitted)~\cite{skowyra2018effective,kim2013topology,weimer2012distributed}. To avoid this type of cyber attack, the defender chooses here periodic topology switching, and preprogram the (repeated) periodic switching sequence into the controlled links, and hence avoids sending the topology information to the cyber layer during the system operation.
\end{itemize}

For our defense strategy based on the periodic topology switching \eqref{eq:psd}, we define the following periodic sequence with length of $l$:
\vspace{-0.20cm}
\begin{align}
\mathbf{L} \triangleq \left\{ \underbrace{\sigma(t_0)}_{\tau_{0}},\underbrace{\sigma(t_1)}_{\tau_{1}}, \ldots, \underbrace{\sigma(t_{l-2})}_{\tau_{l-2}}, \underbrace{\sigma(t_{l-1})}_{\tau_{l-1}} \right\},\label{eq:ss}
\end{align}
\vspace{-0.10cm}
\!\!\!where $\tau_{k}$ denotes the dwell time of the activated topology indexed by $\sigma(t_{k})$, i.e., $\tau_{k} = t_{k+1} - t_{k}$.

Next, we study whether the agents in the system \eqref{eq:ooiio} using control input \eqref{eq:lci} can reach consensus under periodic topology switching. We first recall the well-known property of Laplacian matrix ${\mathcal{L}_r}$ of a connected undirected graph from~\cite{ada11}:
\vspace{-0.60cm}
\begin{subequations}
\begin{align}
{Q_{r}^\top } &= {Q_{r}^{ - 1}},\label{eq:wp1}\\
{\left[ {{Q_r}} \right]_{1,1}} &= {\left[ {{Q_r}} \right]_{2,1}} =  \ldots  = {\left[ {{Q_r}} \right]_{\left| \mathbb{V} \right|,1}}, \label{eq:wp2}\\
{Q_{r}^ \top }\mathcal{L}_{r}Q_{r} &= \emph{\emph{diag}}\left\{ {0,{\lambda _2}(\mathcal{L}_{r}), \ldots , {\lambda _n}(\mathcal{L}_{r})} \right\} \triangleq {\Lambda _r},\label{eq:wp3}
\end{align}\label{eq:ufo4}
\end{subequations}
based on which, we define:
\begin{subequations}
\begin{align}
{\Upsilon _{rs}} &\triangleq Q_r^\top{\mathcal{L}_s}{Q_r},\label{eq:sp1}\\
\mathcal{A}_{s} &\triangleq \left[
    \begin{array}{c;{2pt/2pt}c}
        {{\mathbf{0}_{\left( {\left| \mathbb{V} \right| - 1} \right) \times \left( {\left| \mathbb{V} \right| - 1} \right)}}} & {{\mathbf{1}_{\left( {\left| \mathbb{V} \right| - 1} \right) \times \left( {\left| \mathbb{V} \right| - 1} \right)}}} \\ \hdashline[2pt/2pt]
   {-{\left[ {{\Upsilon _{rs}}} \right]_{{2:\left| \mathbb{V} \right|}, {2:\left| \mathbb{V} \right|}}}}  &  {-{\mathbf{1}_{\left( {\left| \mathbb{V} \right| - 1} \right) \times \left( {\left| \mathbb{V} \right| - 1} \right)}}}\end{array}
\right].\label{eq:sp2}
\end{align}\label{eq:spsp}
\end{subequations}

\begin{prop}
Consider the second-order multi-agent system \eqref{eq:ooiio} with control input \eqref{eq:lci}.  If the sequence $\mathbf{L}$ in \eqref{eq:ss} includes one connected topology, there exists a periodic topology sequence that satisfies
\begin{align}
\sum\limits_{s = 0}^{l-1} {{\nu _s}\mu_{P}\left( {{\mathcal{A}_{s}}} \right)}  < 0,\label{eq:ch1}
\end{align}
where ${\nu _s} = \frac{{{\tau _s}}}{{\tau }}$ with $\tau  = \sum\limits_{i = 0}^{l-1} {{\tau _i}}$. Moreover, under that periodic topology switching, the consensus \eqref{eq:defc} can be achieved. \label{thm:zzaxx}
\end{prop}
\begin{IEEEproof}
See Appendix B.
\end{IEEEproof}
\vspace{-0.00cm}
\begin{rem}
Proposition~\ref{thm:zzaxx} implies that \emph{periodic} topology switching has no constraint on the magnitudes of coupling weights in achieving consensus, i.e., for any coupling weights there exists a feasible periodic topology switching sequence for consensus. This is in sharp contrast with the \emph{arbitrary} topology switching that imposes a strict condition on the magnitudes of coupling weights in achieving consensus \cite{xie2006consensus}.
\end{rem}

\vspace{-0.40cm}
\subsection{System Description}
Under periodic topology switching, the multi-agent system in \eqref{eq:ooiio}, with the control input given by \eqref{eq:lci} and the outputs of monitored agents in $\mathbb{M}$ subject to the attack signal ${g}_{i}(t)$, can be written as
\begin{subequations}
\begin{align}
\!\!\!\!{{\dot{\breve{x}}}_i}\!\left( t \right) &\!=\! {{\breve v}_i}\!\left( t \right) \label{eq:oox1}\\
\!\!\!\!{{\dot{\breve v}}_i}\!\left( t \right) &\!=\!-{{\breve v}_i}\!\left( t \right) + \!\sum\limits_{j \in \mathbb{V}} \!{a_{ij}^{\sigma \left( t \right)}}\!\!\left( {{{\breve x}_j}\!\left( t \right) \!-\! {{\breve x}_i}\!\left( t \right)} \right) \!+\! \left\{ \begin{array}{l}
\hspace{-0.22cm}{{g}}_i\!\left( t \right)\!, i \!\in\! \mathbb{K}\\
\hspace{-0.22cm}0, \hspace{0.10cm} i \!\in\! {\mathbb{V}} \backslash \mathbb{K}
\end{array} \right.\label{eq:oox2asd}\\
\!\!\!\!{{\breve{y}}_i}\!\left( t \right) &\!=\! c_{i1}\breve x_{i}(t) + c_{i2}\breve v_{i}(t) + d_i{{g}}_i\!\left( t \right)\!, i \!\in\! \mathbb{M}\label{eq:oox3}
\end{align}\label{eq:oofn}\end{subequations}
where $c_{i1}$ and $c_{i2}$ are constant coefficients designed by the defender (system operator), while constant coefficient $d_i$ is designed by the attacker.

\begin{rem}
The model in \eqref{eq:oox2asd} with \eqref{eq:oox2} implies that there are two practical approaches to attack the local control inputs: (i) the attacker directly injects the attack signal to the control architectures of misbehaving agents (target agents) in $\mathbb{K}$; (ii) possibly through breaking the encryption algorithm that protects the communication channels  with misbehaving agents, the attacker injects attack signals to the data sent to controller.
\end{rem}

The system in \eqref{eq:oofn} can be equivalently expressed as a switched system under attack:
\begin{subequations}
\begin{align}
&\dot{{{\breve{z}}}}\left( t \right) = A_{{\sigma}(t)}{{\breve{z}}}\left( t \right) + \breve{g}\left( t\right)  \label{eq:s1a}\\
&\breve{y}\left( t \right) = C\breve{z}(t) + D\breve{g}\left( t\right),\label{eq:s1b}
\end{align}\label{eq:s1}\end{subequations}
where we define:
\vspace{-0.15cm}
\begin{subequations}
\begin{align}
\!\!\! \!\!\breve{z}\left( t \right) &\!\triangleq\! \left[
    \begin{array}{c;{1pt/1pt}c;{1pt/1pt}c;{1pt/1pt}c;{1pt/1pt}c;{1pt/1pt}c}
        \!\!\!{{\breve{x}_1}\left( t \right)} \!\!&\! {\ldots} \!\!&\! {\breve{x}_{\left| \mathbb{V} \right|}}\left( t \right) \!\!&\! {\breve{v}_1}\left( t \right) \!\!&\! {\ldots} \!\!&\! {\breve{v}_{\left| \mathbb{V} \right|}}\left( t \right)
    \end{array}\!\!\!
\right]^\top\!,  \label{eq:ssd1}\\
 \!\!\! \!\!A_{\sigma(t)} &\!\triangleq\! \left[
    \begin{array}{c;{2pt/2pt}c}
        \mathbf{0}_{\left| \mathbb{V} \right| \times \left| \mathbb{V} \right|} & \mathbf{1}_{\left| \mathbb{V} \right| \times \left| \mathbb{V} \right|}\\ \hdashline[2pt/2pt]
        -\mathcal{L}_{\sigma(t)} & -\mathbf{1}_{\left| \mathbb{V} \right| \times \left| \mathbb{V} \right|}
    \end{array}
\right],\label{eq:nm0} \\
\! \!\! \!\!C &\!\triangleq\! \left[
    \begin{array}{c;{1pt/1pt}c}
       C_{1} & C_{2}
          \end{array}
\right],\label{eq:nmok}\\
 \!\!\!\!\!{C_j} &\!\triangleq\! \left[
    \begin{array}{c;{1pt/1pt}c}
   \!\!\!\emph{\emph{diag}}\!\left\{ c_{1j}, \ldots , c_{\left| \mathbb{M} \right|j} \right\} \!\!\!&\!\! \mathbf{0}_{\left| \mathbb{M} \right| \times \left( {\left| \mathbb{V} \right| - \left| \mathbb{M} \right|} \right)}\!\!\!\!
    \end{array}\right]\!, j \!=\! 1, 2\label{eq:u1b}\\
    D & \!\triangleq \left[
    \begin{array}{c;{1pt/1pt}c;{1pt/1pt}c}
   \!\!\mathbf{0}_{\left| \mathbb{M} \right| \times {\left| \mathbb{V} \right|} } \!\!\!&\!\! \emph{\emph{diag}}\!\left\{ d_{1}, \ldots , d_{\left| \mathbb{M} \right|} \right\}  \!\!\!&\!\! \mathbf{0}_{\left| \mathbb{M} \right| \times {\left( {\left| \mathbb{V} \right| - \left| \mathbb{M} \right|} \right)}} \!\!\!\!
    \end{array}\right]\!,\label{eq:nmokbb}\\
    \!\!\!\!\!\breve{g}(t) &\!\triangleq\! \left[\!\!
    \begin{array}{c;{1pt/1pt}c}
        {\bf{0}}_{\left| \mathbb{V} \right|}^\top & \bar{g}^{\top}(t)
    \end{array}
\!\!\right]^\top,\label{eq:nm2}\\
\!\!\!\!\!\bar{g}_{i}(t) &\!\triangleq\! \left\{ \begin{array}{l}
\hspace{-0.2cm}{{g}_i}(t),i \in \mathbb{K}\\
\hspace{-0.2cm}0, \hspace{0.52cm}i \in {\mathbb{V}} \backslash \mathbb{K}.
\end{array} \right.\label{eq:u1}
\end{align}
\end{subequations}In addition, we consider the system \eqref{eq:s1} in the absence of attacks, which is given by
\vspace{-0.15cm}
\begin{subequations}
\begin{align}
\dot{{{{z}}}}\left( t \right) &= A_{{{\sigma}}(t)}{{{z}}}\left( t \right),\label{eq:s21}\\
y\left( t \right) &= C{z}(t).\label{eq:s22}
\end{align}\label{eq:s2}\end{subequations}

\vspace{-1.0cm}
\subsection{Privacy of Initial Condition and Global System State}
\vspace{-0.1cm}
To fully  secure multi-agent systems, e.g., connected autonomous vehicles, the initial conditions should be kept confidential from an adversary since the initial data could be utilized to estimate the target locations \cite{xue2014security}. Moreover, individual initial positions contain the information of home-base locations. The following two examples illustrate that the global initial condition as well as the global system state play an important role in stealthy attacks.

\begin{ex}[Attack Objective]
The state solution under attack \eqref{eq:rs2} implies that if $\eta = 0$, attacker's objective is to modify the steady-state value.  If the attack objective is to modify the target location to a new location that the attacker desires, the attacker must know the original target location in the absence of attacks. Under undirected communication, it is straightforward to verify from the system \eqref{eq:ooiio} with its control input \eqref{eq:lci} that the average position $\bar x(t) \buildrel \Delta \over = \frac{1}{{\left| \mathbb{V} \right|}}\sum\limits_{i \in \mathbb{V}} {{x_i}\left( t \right)}$ proceeds with the average velocity $\bar v(t) \buildrel \Delta \over = \frac{1}{{\left| \mathbb{V} \right|}}\sum\limits_{i \in \mathbb{V}} {{v_i}(t)}  = {e^{ - t}}\bar v(t_{0})$, which indicates that when the consensus is achieved, all of the individual agents synchronize to the target location:
\begin{align}
{x^*} = \mathop {\lim }\limits_{t \to \infty } \left( {\bar x(t_{0}) + \left( {1 \!-\! {e^{ - t}}} \right)\bar v(t_{0})} \right){\rm{ }} = \bar x(t_{0}) + \bar v(t_{0}).\label{eq:das}
\end{align}
Unfortunately, \eqref{eq:das} shows that once  the global initial condition is known (i.e., initial positions and velocities of all agents), the original target location can simply be computed through a simple mean computation.
\end{ex}

\begin{ex}[Stealthy Topology Attack Design]
Stealthy topology attack design, as in smart grids~\cite{kim2013topology} and power networks~\cite{weimer2012distributed}, requires  (estimated) real-time data of system states to choose the target connection links to  maliciously alter. Since attacker can record the newly obtained knowledge of the network topology, the attacker has the memory of the past topology sequence. Whenever the data on the global initial condition $z\left( t_{0} \right)$ (or real-time global state $z(t)$) is available, the attacker can infer the exact real-time global state $z(t)$ (or global initial condition $z(t_{0})$) through
\begin{align}
z(t) = {e^{{A_{\sigma \left( {{t_{k}}} \right)}}\left( {t - {t_{k}}} \right)}}\prod\limits_{l = 0}^{k-1} {{e^{{A_{\sigma \left( {{t_l}} \right)}}\left( {{t_{l + 1}} - {t_l}} \right)}}}z\left( t_{0} \right),t \in \left[ {{t_{k}},{t_{k + 1}}} \right) \nonumber
\end{align}
which indicates whenever ZDA seeks a cooperation with stealthy topology attack to evade detection, the attacker would have the largest scope of attackable links since the attacker knows all of agents' real-time state data. Therefore, the private global initial condition or system state can reduce the scope of target links for stealthy topology attack.
\end{ex}

We next impose the following unobservability condition on the monitored outputs to preserve the privacy of non-monitored agents, such that the attacker cannot use the available (monitored) outputs to infer any non-monitored agent's full state (and consequently, the global system state and initial condition).
\begin{lem}\label{lem:switching_observ}
For the system \eqref{eq:s2}, $x_{i}(t)$ and $v_{i}(t)$, $\forall i \in {\mathbb{V}} \backslash \mathbb{M}$, are not simultaneously observable for any $t \in [t_{0}, t^{+}_{m})$, if and only if
\begin{align}
\exists p \in \mathbf{N}^{m}_{0}\!:  \left| p_{i}\right| + \left| p_{i+\left| \mathbb{V} \right|}\right| \neq 0, \forall i \in {\mathbb{V}} \backslash \mathbb{M} \label{eq:ccd}
\end{align}
where \begin{align}
\mathbf{N}^{m}_{m} &= \ker\left ( {\mathcal{O}}_{m} \right ), \label{eq:ck1}\\
\mathbf{N}^{m}_{q} &= \ker({\mathcal{O}}_{q} ) \bigcap e^{-A_{\sigma(t_{q})}\tau_{q}}\mathbf{N}^{m}_{q+1}, 0 \leq q \leq m-1 \label{eq:ck2}\\
{\mathcal{O}}_{q} &= \left[\!\!\!
    \begin{array}{c;{1pt/1pt}c;{1pt/1pt}c;{1pt/1pt}c}
        {{C^\top}} & {{( {C{A_{\sigma(t_{q})}}})^\top}} & \ldots & {{( {CA_{\sigma(t_{q})}^{2\left| \mathbb{V} \right| - 1}})^\top}}
    \end{array}\!\!\!
\right]^\top.  \label{eq:om1}
\end{align}
\end{lem}

\begin{IEEEproof}
The condition in \eqref{eq:ccd} implies that $\mathbf{N}^{m}_{0} \neq \left\{ {{{\bf{0}}_{2\left| \mathbb{V} \right|}}} \right\}$. Using  Theorem 1 in~\cite{tanwani2013observability}, it follows that the system in \eqref{eq:s2} is unobservable for any $t \in [t_{0}, t^{+}_{m})$. Also, \eqref{eq:ccd} implies that $p_{i} \neq 0$, and (or) $p_{i+\left| \mathbb{V} \right|} \neq 0$, and therefore the agent $i$'s position and (or) velocity are (is) not partially observable.
\end{IEEEproof}

\begin{rem}
Although the selection of the monitored output coefficients in \eqref{eq:oox3} subject to \eqref{eq:ccd} renders the system \eqref{eq:s2} unobservable to preserve privacy, we will show that the proposed ZDA variations become detectable using the outputs $y_i(t)$'s by careful selection of switching topologies and the set of monitored agents.
\end{rem}

\section{Stealthy Attack Model}
In the scenario where the attacker is aware of the detection purpose of strategic changes in system dynamics induced by topology switching~\cite{ma1811}, the attacker can evolve the attack policies in response to the strategic changes at switching times to stay stealthy:
\begin{itemize}
  \item ``pause attack" before topology switching when the incoming topology is unknown or the attack policy \eqref{eq:czeo} is infeasible under the known incoming topology, and ``resume attack" after the feasibility of (updated if needed) attack policy under the inferred activated topology is verified;
  \item cooperate with a topology attack that maliciously alters network topology at switching times, such that the original attack policy \eqref{eq:czeo} continues to be feasible under the corrupted topology.
\end{itemize}
In the following subsections, we present a systematic study on these ZDA variations. 

\vspace{-0.40cm}
\subsection{Intermittent Zero-Dynamics Attack}
\vspace{-0.10cm}
For convenience, we refer to $\mathbb{T}$ as the set of topologies under which the attacker injects attack signals to control inputs, and we refer to ${\xi_{k}}$ and $\zeta_{k}$ as the attack-resuming and attack-pausing times over the active topology intervals $[t_{k}, t_{k+1})$, $k \in \mathbb{N}_{0}$, respectively.

The ZDA signals injected into the control input and monitored output of system \eqref{eq:oofn} with intermittent pausing and resuming behaviors are described as
\begin{align}
{{g}_i}( t ) = \left\{ \begin{array}{l}
\hspace{-0.2cm}{g^{\sigma(t_{k})}_{i}}{e^{\eta_{\sigma(t_{k})}( {t - {\xi_{k}}})}}, \hspace{0.2cm}t \!\in\! [ {{\xi_{k}},{{\zeta_{k}}}}) \!\subseteq\! \left[{{t_k},{t_{k + 1}}} \!\right)\\
\hspace{-0.2cm}0,\hspace{2.7cm}\emph{\emph{otherwise}}.
\end{array} \right.\label{eq:ads}
\end{align}

To analyze this ZDA, we review the monitored output \eqref{eq:oox3} at the first ``pausing" time $\zeta_{0}$:
\begin{align}
{\breve{y}_i}\left( \zeta^{-}_{0} \right) = c_{i1}\breve x_{i}(\zeta^{-}_{0}) + c_{i2}\breve v_{i}(\zeta^{-}_{0}) + d_i{{g}}_i\left( \zeta^{-}_{0} \right), \forall i \!\in\! \mathbb{M}\nonumber
\end{align}
which implies that ${\breve{y}_i}\left( \zeta^{-}_{0} \right) = {\breve{y}_i}\left( \zeta_{0} \right)$ if and only if ${{g}_i}\left( \zeta^{-}_{0} \right) = {{g}_i}\left( \zeta_{0} \right)$, since ${\breve{v}_i}\left( \zeta^{-}_{0} \right) = {\breve{v}_i}\left( \zeta_{0} \right)$ and ${\breve{x}_i}\left( \zeta^{-}_{0} \right) = {\breve{x}_i}\left( \zeta_{0} \right)$. Meanwhile, the velocity and position states are always continuous with respect to time, and hence the monitored outputs must be continuous as well. Therefore, to avoid the ``jump" on monitored outputs to maintain the stealthy property \eqref{eq:rs1}, the attacker cannot completely pause the attack, i.e., whenever the attacker pauses injecting ZDA signals to control inputs at pausing time $\zeta_{k}$, simultaneously continues to inject the same attack signals to monitored outputs \eqref{eq:oox3}:
\vspace{-0.20cm}
\begin{align}
\!\!\!{\breve{y}_i}\left( t \right) \!=\! c_{i1}\breve x_{i}(t) \!+\! c_{i2}\breve v_{i}(t) \!+\! d_i\!\!\sum\limits_{m = 0}^{k} \!{{g}_i}\left( \zeta^{-}_{m} \right),t \!\in\! \left[ {\zeta_{k}, {\xi_{k + 1}}} \right)\label{eq:ao1}
\end{align}
or equivalently,
\vspace{-0.20cm}
\begin{align}
{\breve{y}}\left( t \right) = C{\breve{z}}\left( t \right) + D\!\!\sum\limits_{m = 0}^{k} {\breve{g}}\left( \zeta^{-}_{m} \right),t \in \left[ {\zeta_{k}, {\xi_{k + 1}}} \right).\label{eq:ao2}
\end{align}
\vspace{-0.10cm}
Based on the above analysis, for ZDA policy consisting of ``pause attack" and ``resume attack" behaviors to remain stealthy, it should satisfy \eqref{eq:ao2} and
\begin{subequations}
\begin{align}
\mathbf{z}\left( {{t_0}} \right) &\in  \widehat{\mathbf{N}}^{k}_{0} \bigcap \widetilde{\mathbf{N}}^{k}_{0}, \label{eq:ap1}\\
\left[
    \begin{array}{c}
        \mathbf{z}\left( {{{\xi _k}}} \right)  \\ \hdashline[2pt/2pt]
        -\breve{g}\left( {{{\xi_k}}} \right)
    \end{array}
\right] &\in \ker \left( {{\mathcal{P}_r}} \right), \forall \sigma \!\left( {{{\xi_k}}} \right) \!\in\! \mathbb{T}\label{eq:ap2}
\end{align}\label{eq:zkk}
\end{subequations}
where
\begin{align}
\widehat{\mathbf{N}}^{k}_{k} &= \ker ({\mathcal{O}}_{k} ),\label{eq:cm1}\\
\widehat{\mathbf{N}}^{k}_{q} &= \ker({\mathcal{O}}_{q} ) \bigcap e^{-A_{\sigma(t_{q})}(\tau_{q} - \left( {{\zeta _q} - {{\xi_q}}} \right))}\mathbf{N}^{k}_{q+1}, 0 \leq q \leq k \!-\! 1\label{eq:cm2}\\
\widetilde{\mathbf{N}}^{k}_{k} &= \ker ({\widetilde{\mathcal{O}}}_{k} ),\label{eq:cm3}\\
\widetilde{\mathbf{N}}^{k}_{q} &= \ker({\widetilde{\mathcal{O}}}_{q} ) \bigcap e^{-A_{\sigma(t_{q})}(\tau_{q} - \left( {{\zeta _q} - {{\xi_q}}} \right))}\mathbf{N}^{k}_{q+1}, 0 \leq q \leq k \!-\! 1\label{eq:cm4}\\
{\widetilde{\mathcal{O}}}_{r} &\triangleq \left[\!\!\!
    \begin{array}{c;{1pt/1pt}c;{1pt/1pt}c;{1pt/1pt}c}
        {{( {C{A_r}})^\top}} & {{( {C{A^{2}_r}})^\top}} & \ldots & {{( {CA_r^{2\left| \mathbb{V} \right|}} )^\top}}
    \end{array}\!\!\!
\right]^\top,  \label{eq:om1poa}\\
{\mathcal{P}_r} &\triangleq  \left[
    \begin{array}{c;{2pt/2pt}c}
        {\eta_{r}{{\bf{1}}_{2\left| \mathbb{V} \right| \times 2\left| \mathbb{V} \right|}} - {A_{r}}} & {\bf{1}}_{2\left| \mathbb{V} \right| \times 2\left| \mathbb{V} \right|}\\ \hdashline[2pt/2pt]
        - C  & D
    \end{array}
\!\right], \label{eq:pal21} \\
{\mathbf{z}}&= \left[
    \begin{array}{c;{1pt/1pt}c}
        \!\!{\mathbf{x}^\top} \!\!\!&\! {\mathbf{v}^\top}\!\!\!\!
    \end{array}
\right]^\top \triangleq \breve{z} - z = \left[
    \begin{array}{c;{1pt/1pt}c}
        \!\!{\breve{x}^\top} - {x^\top} \!\!\!&\! {\breve{v}^\top} - {v^\top}\!\!\!
    \end{array}
\right]^\top\!\!\!\!,\label{eq:ed0}
\end{align}
and ${\cal O}_{r}$ is given by \eqref{eq:om1}.

\begin{prop}
Under the stealthy attack policy consisting of \eqref{eq:ao2} and \eqref{eq:zkk}, the states and monitored outputs of the systems \eqref{eq:s2} and \eqref{eq:s1} in the presence of attack signal \eqref{eq:ads}  satisfy
\begin{align}
\breve{y}\left( t \right) &= y\left( t \right), t \in \left[ {{{t_0}},{{t_{k+1}}}} \right),\label{eq:cad}\\
\breve{z}\left( t \right) &= z\left( t \right) + {e^{\eta_{\sigma(t_{k)}}\left( {t - {{\xi_k}}} \right)}}{\mathbf{z}\left( {{{\xi_k}}} \right)}, t \in \left[ {{{\xi _k}},{{\zeta_{k}}}} \right).\label{eq:rs2as}
\end{align}\label{thm:ccoo2}
\end{prop}
\vspace{-0.60cm}
\begin{IEEEproof}
See Appendix C.
\end{IEEEproof}

\begin{rem}
At first glance, it might seem that  the intermittent ZDA is an asynchronous attack response to the strategic topology switching, which is due to the imposed \emph{non-negligible} time on capability 3) in Assumption \ref{thm:att}. We note however that the attacker can record the newly obtained topology knowledge into the memory. Since the defender switches topologies {\it periodically}, if the recorded length of topology sequence is sufficiently long, the attacker can learn from the recorded memory the (recurring) periodic sequence, i.e., the attacker knows all future switching topologies and times. The corresponding future synchronous attack policies can be obtained off-line. Therefore, a synchronous attack response is possible only after the attacker obtains the (recurring) periodic topology sequence from memory.
\end{rem}

\vspace{-0.30cm}
\subsection{Cooperative Zero-Dynamics Attack}
The objective of cooperation with stealthy topology attack is to make the ZDA policy \eqref{eq:czeo} continue to hold under the corrupted topology. Stealthy topology attack can be of two types:
\begin{itemize}
  \item Physical Topology Attack: the attacker maliciously alters the status of target connection links of physical systems,  e.g., the bus interaction breaks in power networks~\cite{weimer2012distributed} and link fabrication in software-defined networks~\cite{skowyra2018effective}.
  \item Cyber Topology Attack: the attacker maliciously alters the information of network topology sent to the estimator/observer/detector in cyber layer~\cite{zhang2015implementation,kim2013topology}.
\end{itemize}

As stated in Subsection \ref{III.B}, the basis of our defense strategy is the periodic topology switching, and the defender (system operator) would preprogram the repeated switching times and topologies into the controlled links of the real system and observer/detector. In this case, the operator of the real system does not need to send the topology information to the observer/detector when the system operates. Therefore, the system under our defense strategy is not subject to a cyber topology attack, albeit it is subject  to  a physical topology attack.

We let $t_{k+1}$ denote the switching time when ZDA cooperates with topology attack. The multi-agent system \eqref{eq:s1} in the presence of such cooperative attacks is described by
\begin{subequations}
\begin{align}
\dot{{{\breve{z}}}}\left( t \right) &= \widehat{A}_{{\sigma}(t)}{{\breve{z}}}\left( t \right) + \breve{g}\left( t\right), t \in [t_{k+1},t_{k+2}) \label{eq:s1a1}\\
\breve{y}\left( t \right) &= C\breve{z}(t) + D\breve{g}\left( t\right),\label{eq:s1b1}
\end{align}\label{eq:sas}\end{subequations}
where $\widehat{A}_{{\sigma}(t)}$ is defined as
\begin{equation}\label{eq:A_hat}
    \widehat{A}_{\sigma(t)} \!\triangleq\! \left[
    \begin{array}{c;{2pt/2pt}c}
        \mathbf{0}_{\left| \mathbb{V} \right| \times \left| \mathbb{V} \right|} & \mathbf{1}_{\left| \mathbb{V} \right| \times \left| \mathbb{V} \right|}\\ \hdashline[2pt/2pt]
        -\widehat{\mathcal{L}}_{\sigma(t)} & -\mathbf{1}_{\left| \mathbb{V} \right| \times \left| \mathbb{V} \right|}
    \end{array}
\right],
\end{equation}
with ${{\widehat{\mathcal{L}}}_{\sigma \left( t_{k + 1} \right)}}$ denoting the Laplacian matrix of the corrupted topology.
We describe its corresponding system in the absence of ZDA, i.e., in the presence of the only physical topology attack, as
\begin{subequations}\label{eq:sasb}
\begin{align}
\dot{\widehat{{z}}}\left( t \right) &= {{\widehat A}_{\sigma (t)}}\widehat{{z}}\left( t \right),t \in \left[ {{t_{k + 1}},{t_{k + 2}}} \right)\\
\widehat{{y}}\left( t\right) &= C\widehat{{z}}\left( t \right).
\end{align}\end{subequations}

If $\breve{g}\left( t\right)$ is a ZDA signal in systems \eqref{eq:s1} and \eqref{eq:sas} at times $t_{k + 1}^ -$ and $t_{k + 1}$,  by \eqref{eq:rs2} we have
$\breve{z}\left( t_{k + 1} \right) = \breve{z}\left( t^{-}_{k + 1} \right) = {{z}}\left( t^{-}_{k + 1} \right) + {\mathbf{z}}_{0}{e^{\eta t_{k + 1}}}$ and $ \breve{z}\left( t^{-}_{k + 1} \right)= \breve{z}\left( t_{k + 1} \right) =  \widehat{{z}}\left( t_{k + 1} \right) + {\mathbf{z}}_{0}{e^{\eta t_{k + 1}}}$. Here, we conclude that
\begin{align}
\widehat{{z}}\left( {{t_{k + 1}}} \right) = z\left( {{t^{-}_{k + 1}}} \right) = z\left( {{t_{k + 1}}} \right) ,\label{eq:kkc}
\end{align}
otherwise, the system state $\breve{z}\left( t_{k + 1} \right)$ has ``jump" behavior, which contradicts with the fact that $\breve{z}(\cdot)$ is continuous.

The equation \eqref{eq:kkc} and the stealthy property \eqref{eq:rs1} imply that $C\breve{z}\left( {{t_{k + 1}}} \right) = Cz\left( {{t_{k + 1}}} \right) = C\widehat{{z}}\left( {{t_{k + 1}}} \right)$, based on which, a necessary condition for the existence of ZDA under corrupted topology is stated formally in the following proposition.

\begin{prop}
Consider the systems in \eqref{eq:sasb} and \eqref{eq:s2}.  We have $y\left( t \right)$ = $\widehat{{y}}\left( {{t}} \right)$ for any $t \in [t_{k+1},t_{k+2})$, if and only if
\begin{align}
&\sum\limits_{l = 0}^d {C\widehat A_{\sigma \left( {{t_{k + 1}}} \right)}^l({{\widehat A}_{\sigma \left( {{t_{k + 1}}} \right)}} - {A_{\sigma \left( {{t_{k + 1}}} \right)}}){z^{(d - l)}}\left( {{t_{k + 1}}} \right)  }        \nonumber\\
&\hspace{5.2cm} = {{\bf{0}}_{\left| \mathbb{M} \right|}},\ \forall d \in {\mathbb{N}_0}.  \label{eq:xxz}
\end{align}
\label{thm:nnq}
\end{prop}
\vspace{-0.50cm}
\begin{IEEEproof}
See Appendix D.
\end{IEEEproof}

We set $d= 0, 1$ and expand \eqref{eq:xxz} out to obtain:
\begin{subequations}
\begin{align}
{C_2}(\widehat{\mathcal{L}}_{\sigma( t_{k + 1})} - {\cal L}_{\sigma( {{t_{k + 1}}})})x( {{t_{k + 1}}}) &= {{\bf{0}}_{\left| \mathbb{M} \right|}},\label{eq:xxza11}\\
{C_2}(\widehat{\mathcal{L}}_{\sigma( t_{k + 1})} - {\cal L}_{\sigma( {{t_{k + 1}}})})v( {{t_{k + 1}}}) &= {{\bf{0}}_{\left| \mathbb{M} \right|}}.\label{eq:xxza21}
\end{align}\label{eq:op1}\end{subequations}
The result \eqref{eq:op1} shows that like the stealthy topology attacks in smart grids~\cite{zhang2015implementation,kim2013topology} and software-defined networks~\cite{skowyra2018effective}, the attacker needs some agents' real-time state data to decide  the target links to attack, while according to Lemma \ref{lem:switching_observ}, the attacker cannot simultaneously infer $x_{i}\left( {{t_{k + 1}}} \right)$ and $v_{i}\left( {{t_{k + 1}}} \right)$, $\forall i \in {\mathbb{V}} \backslash \mathbb{M}$. Therefore, there should be a scope of attackable connection links under the strategy \eqref{eq:ccd}.

Without loss of generality, we express the difference of Laplacian matrices in the form:
\begin{align}
\!\!\!{\widehat {\cal L}_{\sigma \left(\! {{t_{k + 1}}} \!\right)}} \!-\! {\mathcal{L}_{\sigma \left(\! {{t_{k + 1}}} \!\right)}} &\!=\!\! \left[
    \begin{array}{c;{2pt/2pt}c}
        \!\!\!\mathfrak{L}_{\sigma(t_{k+1})} \!\!\!&\!\! {{\mathbf{0}_{\left| \mathbb{D} \right| \!\times\! \left( {\left| \mathbb{V} \right| - \left| \mathbb{D} \right|} \right)}}} \!\!\!\!\\ \hdashline[2pt/2pt]
        \!\!\!{\mathbf{0}_{\left( {\left| \mathbb{V} \right| - \left| \mathbb{D} \right|} \right) \!\times\! \left| \mathbb{D} \right|}}  \!\!\!&\!\! {\mathbf{0}_{\left( {\left| \mathbb{V} \right| - \left| \mathbb{D} \right|} \right) \!\times\! \left( {\left| \mathbb{V} \right| - \left| \mathbb{D} \right|} \right)}}\!\!\!\!
    \end{array}
\right],\label{eq:sep}
\end{align}
where $\mathbb{D}$ denotes  the set of agents in the sub-graph formed by the target links to be possibly attacked, $\mathfrak{L}_{\sigma(t_{k+1})} \in \mathbb{R}^{\left| \mathbb{D} \right| \times \left| \mathbb{D} \right|}$ is the elementary row transformation of the Laplacian matrix of a subgraph $\mathcal{G}$ in the difference graph, which is generated by the corrupted graph $\widehat{\mathcal{G}}_{t_{k+1}}$ of the topology attacker and candidate graph ${\mathcal{G}}_{t_{k+1}}$ of the defender at time $t_{k+1}$.

Since $C_{2} \in \mathbb{R}^{\left| \mathbb{M} \right| \times \left| \mathbb{V} \right|}$ and $\mathfrak{L}_{\sigma(t_{k+1})} \in \mathbb{R}^{\left| \mathbb{D} \right| \times \left| \mathbb{D} \right|}$, the relations in \eqref{eq:ccd}, \eqref{eq:op1}, and \eqref{eq:sep} imply that  the attacker can devise a stealthy topology attack (without knowing the measurements of the agents in $\mathbb{V} \backslash \mathbb{M}$ which are unavailable) only when the scope of target links satisfies:
\begin{align}
\mathbb{D} &\subseteq \mathbb{M}. \label{eq:stcl}
\end{align}

\vspace{-0.30cm}
\section{Detectability of Stealthy Attacks}
\vspace{-0.10cm}
Based on the systematic study of the attack behaviors and policies in Section IV,  in this section, we investigate the detectability of the proposed ZDA variations.
\vspace{-0.45cm}
\subsection{Detectability of Intermittent Zero-Dynamics Attack}
\vspace{-0.10cm}
We first define
\begin{align}
{\mathcal{U}_{ri}} &\triangleq \text{diag}\left\{ {{{\left[ {{Q_r}} \right]}_{i,1}}, \ldots ,{{\left[ {{Q_r}} \right]}_{i,\left| \mathbb{V} \right|}}} \right\}{Q^{\top}_r},\label{eq:da}\\
\mathbb{F} &\triangleq \left\{ {\left. i \right|}\left[ {{Q_r}} \right]_{i,j} \ne 0, i \in \mathbb{M}, \forall j \in \mathbb{V}, \forall r \in \mathbf{L} \right\},\label{eq:dahh}
\end{align}
where ${Q_r}$ satisfies \eqref{eq:ufo4}.
\begin{empheq}[box={\Garybox[\small Defense Strategy Against Intermittent ZDA]}]{align}
&\text{Strategy on switching topologies:} ~\mathcal{L}_{r}~\text{has distinct} \nonumber\\
 &\hspace{4.1cm} \text{eigenvalues for}~ \forall r \in \mathbf{L}. \label{eq:kkz0}\\
 &\text{Strategy on monitored-agent locations:} ~\mathbb{F} \ne \emptyset.\label{eq:kkz1}
\end{empheq}
\vspace{-0.20cm}
\begin{thm}
Consider the system \eqref{eq:oofn} in the presence of attack signals \eqref{eq:ads}. Under the defense strategy against intermittent ZDA,
\begin{itemize}
  \item if the monitored agents output the full observations of their velocities (i.e., ${c_{i1}} = 0 ~ \text{and} ~ {c_{i2}} \ne 0$ for $\forall i \in \mathbb{M}$), the intermittent ZDA is detectable and
      \vspace{-0.20cm}
      \begin{align}
 \mathbf{N}^{\infty}_{0} = \left\{ {{{\mathbf{0}}_{2\left| \mathbb V \right|}}}, \left[
    \begin{array}{c;{1pt/1pt}c}
       \!\!\!\mathbf{1}_{\left| \mathbb{V} \right|}^\top \!\!\!&\!\! \mathbf{0}_{\left| \mathbb{V} \right|}^\top \!\!\!
          \end{array}
\right]^\top \right\};\label{eq:oa1}
\end{align}
  \item if the monitored agents output the full observations of their positions (i.e., ${c_{i1}} \ne 0 ~ \text{and} ~ {c_{i2}} = 0$ for $\forall i \in \mathbb{M}$), the intermittent ZDA is detectable but
      \vspace{-0.20cm}
      \begin{align}
\mathbf{N}^{\infty}_{0} = \left\{ {{{\mathbf{0}}_{2\left| \mathbb V \right|}}} \right\};\label{eq:oa2}
\end{align}
\vspace{-0.60cm}
  \item if the monitored agents output the partial observations (i.e., ${c_{i1}} \ne 0 \hspace{0.1cm} \text{and} \hspace{0.1cm} {c_{i2}} \ne 0$ for $\forall i \in \mathbb{M}$), and $c_{i1} = c_{i2}, \forall i \in \mathbb{M}$, the kernel of the observability matrix satisfies
      \vspace{-0.20cm}
  \begin{align}
 \mathbf{N}^{\infty}_{0} = \left\{ {{{\mathbf{0}}_{2\left| \mathbb V \right|}}}, \left[
    \begin{array}{c;{1pt/1pt}c}
       \!\!\!\mathbf{1}_{\left| \mathbb{V} \right|}^\top \!\!\!&\!\! -\mathbf{1}_{\left| \mathbb{V} \right|}^\top \!\!\!
          \end{array}
\right]^\top \right\};\label{eq:oa1k}
\end{align}
and the intermittent ZDA is detectable if
\vspace{-0.10cm}
\begin{align}
\xi_{0} > t_{0}, ~\text{or}~ D = {\mathbf{0}_{\left| \mathbb{M} \right| \times 2\left| \mathbb{V} \right|}}, \label{eq:nad1}
\end{align}
where ${\mathbf{N}}^\infty_{0}$ is computed recursively by~(\ref{eq:ck1}) and~(\ref{eq:ck2}).
\end{itemize}
\label{thm:thd}
\end{thm}
\begin{IEEEproof}
See Appendix E.
\end{IEEEproof}

Under the defense strategy consisting of \eqref{eq:kkz0} and \eqref{eq:kkz1}, the result \eqref{eq:oa2} implies that if the monitored agents output full observations of position,  the condition \eqref{eq:ccd}  is not satisfied. While the results \eqref{eq:oa1} and \eqref{eq:oa1k} show that if the monitored agents output full observations of velocity or partial observations, the condition \eqref{eq:ccd} is satisfied, and according to Lemma \ref{lem:switching_observ}, the privacy of all states of non-monitored agents is preserved, which further implies that using the available data \eqref{eq:rs1}, the attacker cannot infer the global system state and the global initial condition. Therefore, for the purpose of privacy preserving of non-monitored agents' states, consequently, restricting the scope of attackable links to derive the defense strategies against the cooperative ZDA, we abandon full observation of position.

\vspace{-0.10cm}
\subsection{Detectability of Cooperative Zero-Dynamics Attack}
\vspace{-0.00cm}
Considering the matrix ${Q_r}$ satisfying \eqref{eq:ufo4}, we describe the defense strategy as follows:
\begin{empheq}[box={\Garybox[\small Defense Strategy Against Cooperative ZDA]}]{align}
& \text{Strategy on switching topologies:} ~(\ref{eq:kkz0}). \nonumber\\
 & \text{Strategy on monitored-agent outputs:} ~c_{i2} \!>\! 0, \forall i \!\in\! \mathbb{M}. \label{eq:dfv}\\
 & \text{Strategy on monitored-agent locations:} \nonumber\\
 &\small{{[{Q_r}]}_{i,m}} \!\!-\! {{[ {{Q_r}}]}_{j,m}} \!\!\neq\! 0, \forall m \!\in\! {\mathbb{V}} \backslash \{1\}, \forall r \!\in\! \mathbf{L},\forall i \!\neq\! j \!\in\! \mathbb{M}. \label{eq:kkz1x}
\end{empheq}

\begin{thm}
Consider the system \eqref{eq:sas} in the presence of zero-dynamics attack in cooperation with topology attack under \eqref{eq:stcl}. Under the defense strategy against cooperative ZDA, the attack is detectable. \label{thm:thdb}
\end{thm}
\begin{IEEEproof}
See Appendix F.
\end{IEEEproof}

\begin{rem}
The common critical requirement of our defense strategies is that the communication network has distinct Laplacian eigenvalues. There indeed exist many topologies whose associated Laplacian matrices have distinct eigenvalues. The following lemma provides a guide to design such topologies:
\begin{lem}[Proposition 1.3.3 in~\cite{ada11}]
Let $\mathrm{G}$ be a connected graph with diameter $m$. Then, $\mathrm{G}$ has at least $m + 1$ distinct Laplace eigenvalues.\label{thm:paa}
\end{lem}
\end{rem}

\vspace{-0.20cm}
\section{Attack Detection Algorithm}
\vspace{-0.00cm}
Using the proposed defense strategies and the detectability conditions in Section V, this section focuses on the attack detection algorithm that is based on a Luenberger observer.

\vspace{-0.7cm}
\subsection{Luenberger Observer under Switching Topology}
\vspace{-0.10cm}
We now present a Luenberger observer~\cite{lbo}:
\begin{subequations}
\begin{align}
\!\!\!\!\!{q_i}\left( t \right) &= {{w}_i}\left( t \right) \label{eq:fl1}\\
\!\!\!\!\!{{\dot{w}}_i}\!\left( t \right) &= -{{w}_i}\left( t \right) + \sum\limits_{i \in \mathbb{V}} \!{a_{ij}^{\sigma \left( t \right)}}\!\!\left( {{{q}_j}\!\left( t \right) - {{q}_i}\!\left( t \right)} \right) \nonumber\\
&\hspace{1.5cm}- \left\{ \begin{array}{l}
\hspace{-0.2cm}r_{i}(t), \hspace{0.92cm}c_{i1} \neq 0, i \!\in\! \mathbb{M}\\
\hspace{-0.2cm}\int_{t_{0}}^{t}\!{r_{i}(b)}\mathrm{d}b,\hspace{0.1cm} c_{i1} = 0, \hspace{0.00cm} i \!\in\!  \mathbb{M}\\
\hspace{-0.2cm}0, \hspace{1.4cm} i \!\in\! {\mathbb{V}} \backslash \mathbb{M}\\
\end{array} \right. \label{eq:fl2}\\
\!\!\!\!\!{r}_i\!\left( t \right) &= c_{i1} q_{i}(t)  + c_{i2} w_{i}(t) \!-\! \breve{y}_{i}(t) , i \!\in\! \mathbb{M}\label{eq:fl3}
\end{align}\label{eq:fl}\end{subequations}
where $\breve{y}_{i}(t)$ is the monitored output of agent $i$ in system \eqref{eq:oofn}, $r_{i}\left( t  \right)$ is the attack-detection signal.

We next consider a system matrix related to the system \eqref{eq:fl} in the absence of attacks:
\begin{align}
\widehat{\mathcal{A}}_{r} \triangleq \left[
    \begin{array}{c;{1.0pt/1.0pt}c}
        {\mathbf{0}_{\left| \mathbb{V} \right| \times \left| \mathbb{V} \right|}} & {\mathbf{1}_{\left| \mathbb{V} \right| \times \left| \mathbb{V} \right|}}  \\ \hdashline[1pt/1pt]
        { - \mathcal{L}_{r} - {\widehat C}} & -\mathbf{1}_{\left| \mathbb{V} \right| \times \left| \mathbb{V} \right|}
    \end{array}
\right],\label{eq:dm}
\end{align}
where
\begin{align}
\widehat{C} \triangleq \left[
    \begin{array}{c}
       C_{1}   \\ \hdashline[1pt/1pt]
        {\mathbf{0}_{(\left| \mathbb{V} \right| - \left| \mathbb{M} \right|) \times \left| \mathbb{V} \right|}}
    \end{array}
\right] \hspace{0.1cm} \emph{\emph{or}} \hspace{0.1cm}  \left[
    \begin{array}{c}
       C_{2}   \\ \hdashline[1pt/1pt]
        {\mathbf{0}_{(\left| \mathbb{V} \right| - \left| \mathbb{M} \right|) \times \left| \mathbb{V} \right|}}
    \end{array}
\right]  \label{eq:def_cmatrix}
\end{align}
with $C_{1}$ and $C_{2}$ given by \eqref{eq:u1b}. It is straightforward to obtain the following result regarding the matrix stability.
\begin{lem}
The matrix $\widehat{\mathcal{A}}_{r}$ defined by \eqref{eq:dm} is Hurwitz, if $\mathcal{L}_{r}$ is the Laplacian matrix of a connected graph and
\begin{align}
\mathbf{0}_{\left| \mathbb{V} \right| \times \left| \mathbb{V} \right|} \neq {\widehat C} \geq 0.\label{eq:cog}
\end{align} \label{thm:my0bd}
\end{lem}

\vspace{-0.50cm}
If the sequence \eqref{eq:ss} has one connected graph and gain matrix $\widehat{C}$ \eqref{eq:def_cmatrix} satisfies \eqref{eq:cog}, it follows from Lemma \ref{thm:my0bd} that there exists a $P > 0$, such that under convex linear combination, the matrix measure satisfies
\begin{align}
\sum\limits_{s = 0}^{l-1} {{\nu _s}\mu_{P}\left( {{\widehat{\mathcal{A}}_{s}}} \right)}  < 0.\label{eq:sdsbk10}
\end{align}

\subsection{Strategic Topology-Switching Algorithm}
We next propose Algorithm~1 that describes when and which topology to switch to detect the ZDA variations.

\begin{algorithm}[http]
  \caption{Strategic Topology Switching}
  \KwIn{Initial index $k$ = 0, initial time $t_{k} = 0$, observer gains satisfying \eqref{eq:cog}, periodic sequence $\mathbf{L}$ \eqref{eq:ss}  with length of $l$ satisfying \eqref{eq:ch1} and \eqref{eq:sdsbk10}.}
  Run the system \eqref{eq:oofn} and the observer \eqref{eq:fl}\;
  Update dwell time: ${\tau _{\sigma(t_k)}} \leftarrow {\tau _{\sigma( {{t_{\bmod ( {k,L + 1})}}})}}$\;
  Switch topology of system \eqref{eq:oofn} and observer \eqref{eq:fl} at time $t_{k} + \tau_{{\sigma}(t_{k})}$:
  $\sigma(t_{k} + \tau_{{\sigma}(t_{k})}) \leftarrow \mathbf{L}( {\bmod ( {k + 1,L})})$\;
  Update switching time: $t_{k} \leftarrow t_{k} + \tau_{{\sigma}(t_{k})}$\;
  Update index: $k \leftarrow k+1$\;
  Go to Step 2.
\end{algorithm}

\begin{thm}If the monitored agents satisfy \eqref{eq:kkz1}, \eqref{eq:dfv} and \eqref{eq:kkz1x}, and the switching topologies in $\mathbf{L}$ satisfy \eqref{eq:kkz0},
\begin{itemize}
  \item without requiring the knowledge of the misbehaving agents and the start, pause, and resume times of the attack,
\begin{enumerate}
\item with $c_{i1} = 0$, $\forall i \in \mathbb{M}$, the observer \eqref{eq:fl} is able to detect the intermittent and cooperative ZDAs;
\item with $c_{i1} = c_{i2}$, $\forall i \in \mathbb{M}$, the observer \eqref{eq:fl} is able to detect the cooperative
ZDA and intermittent ZDA under \eqref{eq:nad1};
\end{enumerate}
   \item in the absence of attacks, the agents in system \eqref{eq:oofn} achieve the asymptotic consensus, and the observer \eqref{eq:fl} asymptotically tracks the real system \eqref{eq:s1} if $c_{i1} = c_{i2}$, $\forall i \in \mathbb{M}$, or $c_{i1} = 0$, $\forall i \in \mathbb{M}$.
\end{itemize}
\label{thm:dfd}
\end{thm}
\begin{IEEEproof}
See Appendix G.
\end{IEEEproof}

\vspace{-0.00cm}
\begin{rem}
The modulo operations in steps 2 and 3 of Algorithm~1 describe the building block of our defense strategy, that is periodic topology switching. Given the length of topology switching sequence, i.e., $l$, and the length of the running time of the system \eqref{eq:oofn} and the observer \eqref{eq:fl}, denoted by $t_{f} - t_{0}$, the total number of topology switchings can roughly be computed as $\frac{{{t_f} - {t_0}}}{\tau }l$.
\end{rem}

\section{Simulations}
We consider a system with $n = 16$ agents. The initial position and velocity conditions are chosen as ${x}(t_{0}) = {\left[ {2\times\mathbf{1}^{\top}_{8},4\times\mathbf{1}^{\top}_{8}} \right]^ \top }$ and ${v}(t_{0}) = {\left[ {6\times\mathbf{1}^{\top}_{8},8\times\mathbf{1}^{\top}_{8}} \right]^ \top }$.
The coupling weights and observer gains are uniformly set to one. The considered network topologies are given in the following Figures~\ref{fig:tpi} and~\ref{fig:tpc} where the yellow nodes denote the monitored agents that output full observations of individual velocities.

\vspace{-0.40cm}
\subsection{Detection of Intermittent ZDA}
\vspace{-0.00cm}
\begin{figure}[http]
\centering{
\includegraphics[scale=0.515]{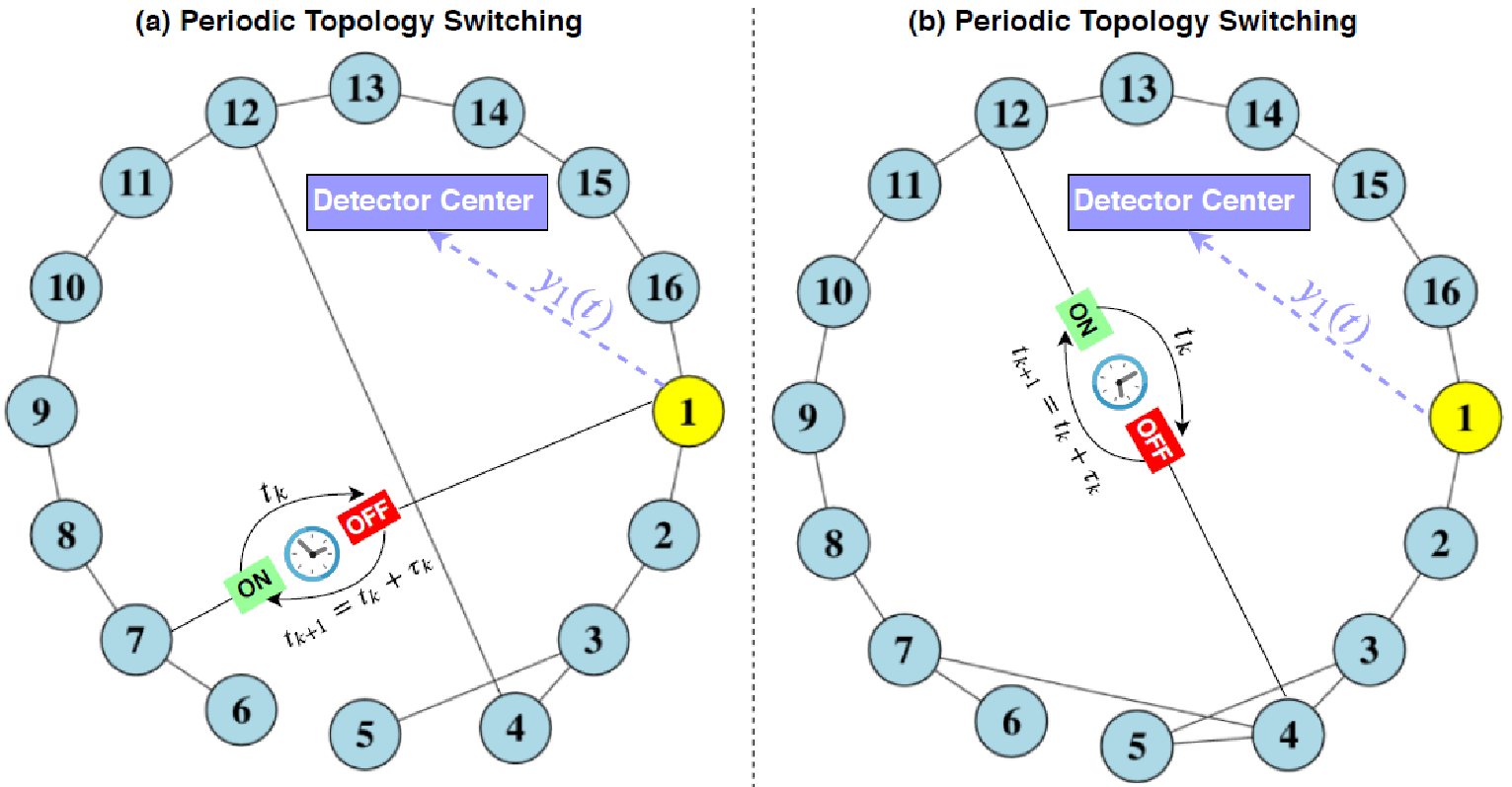}
}
\caption{Two periodic topology switching schemes for intermittent ZDA.}
\label{fig:tpi}
\end{figure}
\vspace{-0.0cm}

We first consider the periodic topology switching scheme in Figure~\ref{fig:tpi} (a). We denote the topologies with the controlled links $a^{\sigma(t)}_{17}$ in ``On" and ``Off" by 1 and 2, respectively. The considered corresponding
periodic switching sequence is $\mathbf{L} = \left\{\underbrace{\sigma(t_0) = 1}_{\tau_{0} = 3}, \underbrace{\sigma(t_1) = 2}_{\tau_{1} = 6}\right\}.$
It can be verified that with $y_{1}(t) = v_{1}(t)$, neither of the switching topologies in Figure~\ref{fig:tpi} (a) has distinct eigenvalues and $\mathbb{F} = \varnothing$, such that the defense strategy consisting of \eqref{eq:kkz0} and \eqref{eq:kkz1} does not hold. Therefore, the attacker can design an undetectable intermittent ZDA as follows:
\begin{itemize}
  \item inject false data $\mathbf{z}(t_{0}) = \left[ {\mathbf{0}^{\top}_{3},-1,1,\mathbf{0}^{\top}_{14},-0.08 - 2\mathrm{i},} \right.\\
  \left. {0.08 + 2\mathrm{i},\mathbf{0}^{\top}_{11}} \right]^\top$ to the data of initial condition sent to the observer \eqref{eq:fl};
  \item inject ZDA signals ${\breve{g}_4}(t)$ $=$ $(2.9136 + 2.32\mathrm{i}){e^{(0.08 - 2\mathrm{i})(t-0.2)}}$ and ${\breve{g}_5}(t)$ $=$ $(-2.9136 - 2.32\mathrm{i}){e^{(0.08 - 2\mathrm{i})(t-0.2)}}$ to the local control inputs of agents 4 and 5 for the initial Topology 1 at $\xi_{0} = 0.2$;
  \item pause the ZDA if the incoming topology is unknown;
  \item update the attack strategy if necessary, and resume the feasible attack after newly switched topology is inferred;
  \item iterate the last two steps.
\end{itemize}

\begin{figure}[http]
\centering{
\includegraphics[height=1.75in,width=3.50in]{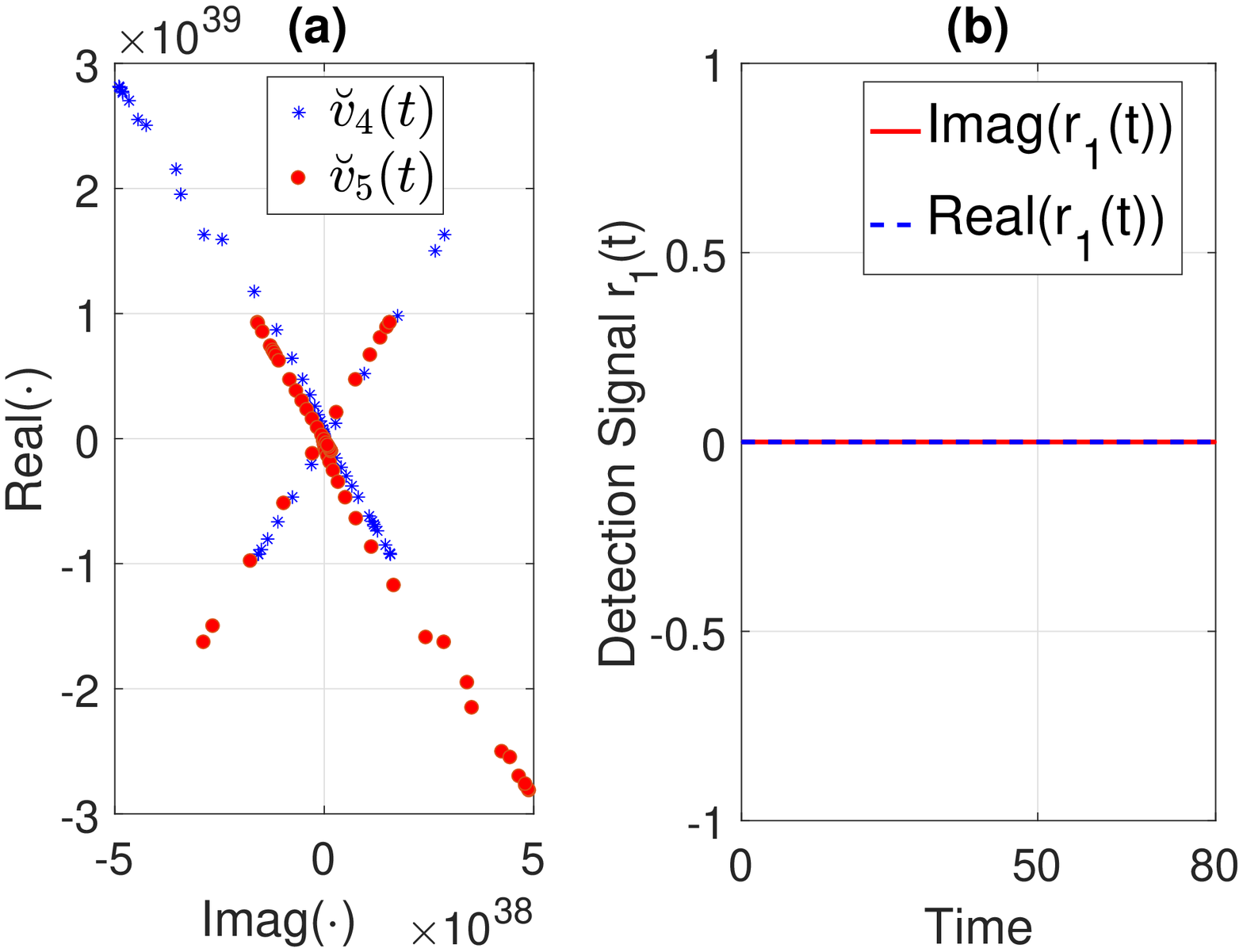}}
\caption{Individual velocities (a) and attack-detection signal (b).}
\label{fig:stiz}
\end{figure}

Some agents' velocities and the attack-detection signals in Figure~\ref{fig:stiz} show that with $y_{1}(t) = v_{1}(t)$, when the defense strategy consisting of \eqref{eq:kkz0} and \eqref{eq:kkz1} does not hold, the attacker can design an intermittent ZDA that cannot be detected by the observer \eqref{eq:fl} under Algorithm~1 (constant zero detection signal), and the stealthy attack renders the system unstable (in the absence of attacks, $\mathop {\lim }\limits_{t \to \infty } | {{v_i}(t)}| = 0, \forall i \in \mathbb{V}$).

\begin{figure}[http]
\centering{
\includegraphics[height=1.70in,width=3.60in]{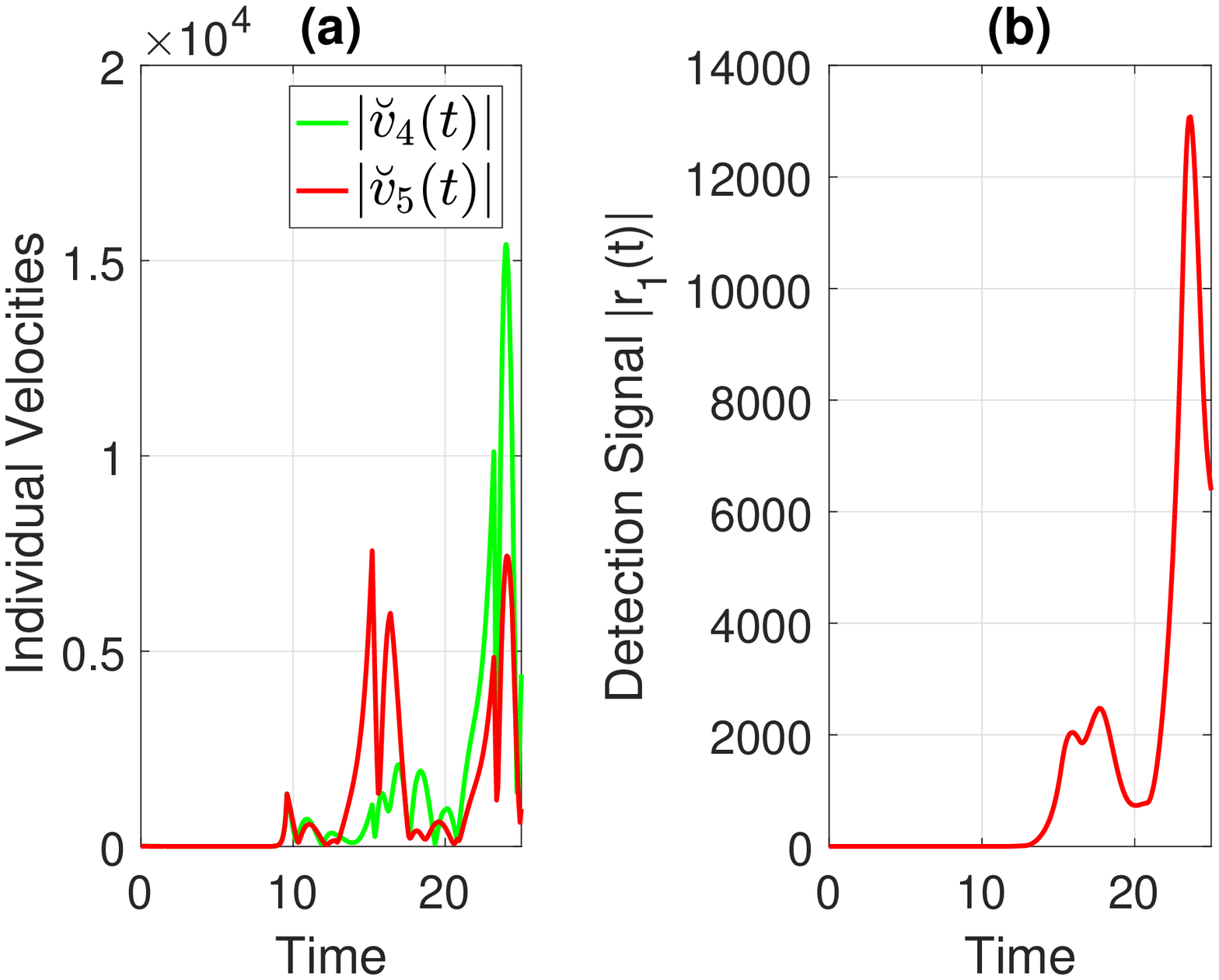}}
\caption{Trajectories of velocities (a) and attack-detection signal (b).}
\label{fig:stizs}
\end{figure}

The switching topologies in Figure~\ref{fig:tpi} (b) satisfy the defense strategy consisting of \eqref{eq:kkz0} and \eqref{eq:kkz1}. Hence, according to the first statement in Theorem \ref{thm:thd}, with $y_{1}(t) = v_{1}(t)$, i.e., $c_{11} = 0$, we can turn to the switching scheme at some time to detect the stealthy attack. Under the topology switching scheme in Figure~\ref{fig:tpi} (b), the trajectory of attack-detection signal in Figure~\ref{fig:stizs} (b) shows the observer \eqref{eq:fl}  succeeds in detecting the intermittent ZDA (nonzero detection signal), which also demonstrates the first statement in Theorem \ref{thm:dfd}.

\vspace{-0.20cm}
\subsection{Detection of Cooperative ZDA}
\vspace{-0.20cm}

\begin{figure}[http]
\centering{
\includegraphics[scale=0.55]{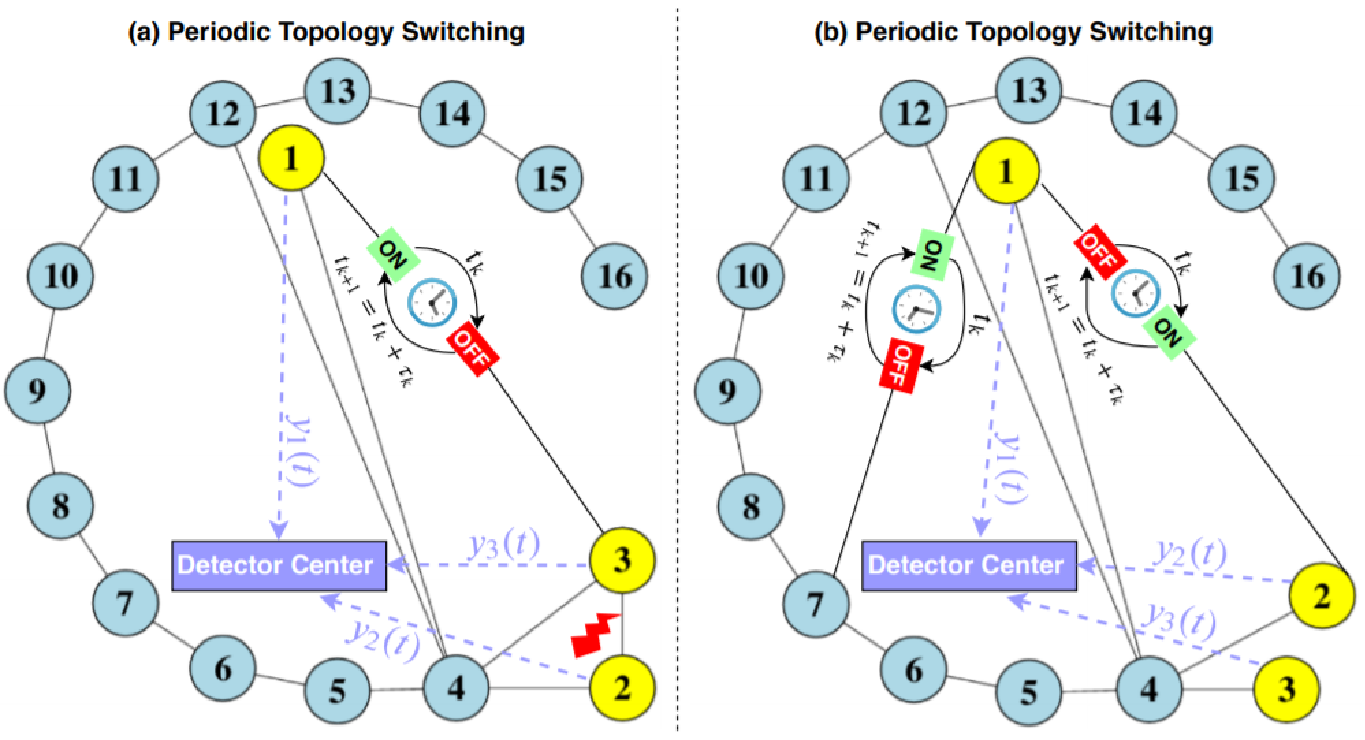}}
\caption{Network topologies for cooperative ZDA.}
\label{fig:tpc}
\end{figure}

\begin{figure}[http]
\centering{
\includegraphics[height=1.70in,width=3.50in]{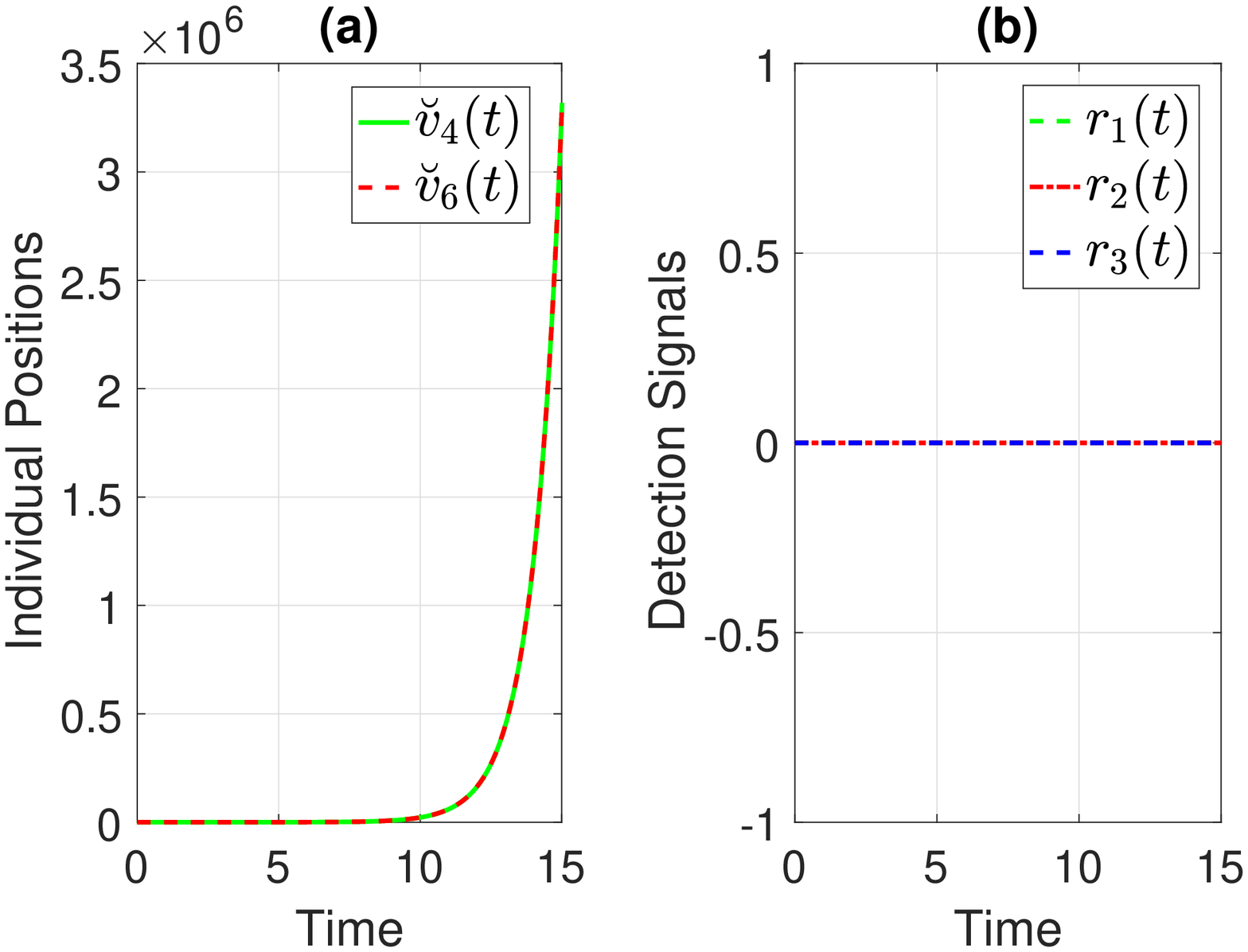}}
\caption{Trajectories of velocities (a) and attack-detection signals (b).}
\label{fig:tpdc}
\end{figure}

We denote the switching topologies in Figure~\ref{fig:tpc} (a) by 3 and 4, in Figure~\ref{fig:tpc} (b) by 5 and 6, respectively. It can be verified that with $y_{i}(t) = v_{i}(t)$, $i \in \mathbb{M} = \{1,2,3\}$, neither Topology 3 nor Topology 4 satisfies the defense strategy consisting of \eqref{eq:kkz0}, \eqref{eq:dfv} and \eqref{eq:kkz1x}. Therefore, under the periodic topology switching sequence
$\mathbf{L} = \left\{\underbrace{\sigma(t_0) = 3}_{\tau_{0} = 3}, \underbrace{\sigma(t_1) = 4}_{\tau_{1} = 1} \right\}$, it is possible to design stealthy cooperative ZDA as follows:
\begin{itemize}
  \item inject false data $\mathbf{z}(t_{0}) = \left[ {0,0,0,1,1,1,0,0,0,0,0,0,0,} \right.\\
  \left. {0,0,0, 0,0,0,1,1,1,0,0,0,0,0,0,0,0,0,0} \right]^\top$ to the data of initial condition sent to observer \eqref{eq:fl};
  \item inject ZDA signals ${\breve{g}_1}\left( t \right) = {\breve{g}_3}\left( t \right) = {\breve{g}_7}\left( t \right) = -{e^{t}}$, ${\breve{g}_4}\left( t \right) = 5{e^{t}}$, ${\breve{g}_5}\left( t \right) = 2{e^{t}}$ and ${\breve{g}_6}\left( t \right) = 3{e^{t}}$ to the local control inputs of agents 1, 3, 7, 4, 5 and 6, respectively, at initial time for Topology 3;
  \item inject false data $-{e^{t}}$ to the monitored outputs;
  \item maliciously control the connection between agents 2 and 3, such that the original ZDA policy maintains its feasibility under the corrupted topology at incoming switching times.
\end{itemize}
The trajectories of velocities and attack-detection signals in Figure~\ref{fig:tpdc} show that the designed attack makes system unstable without being detected (constant zero detection signals).

\begin{figure}[http]
\centering{
\includegraphics[height=1.80in,width=3.50in]{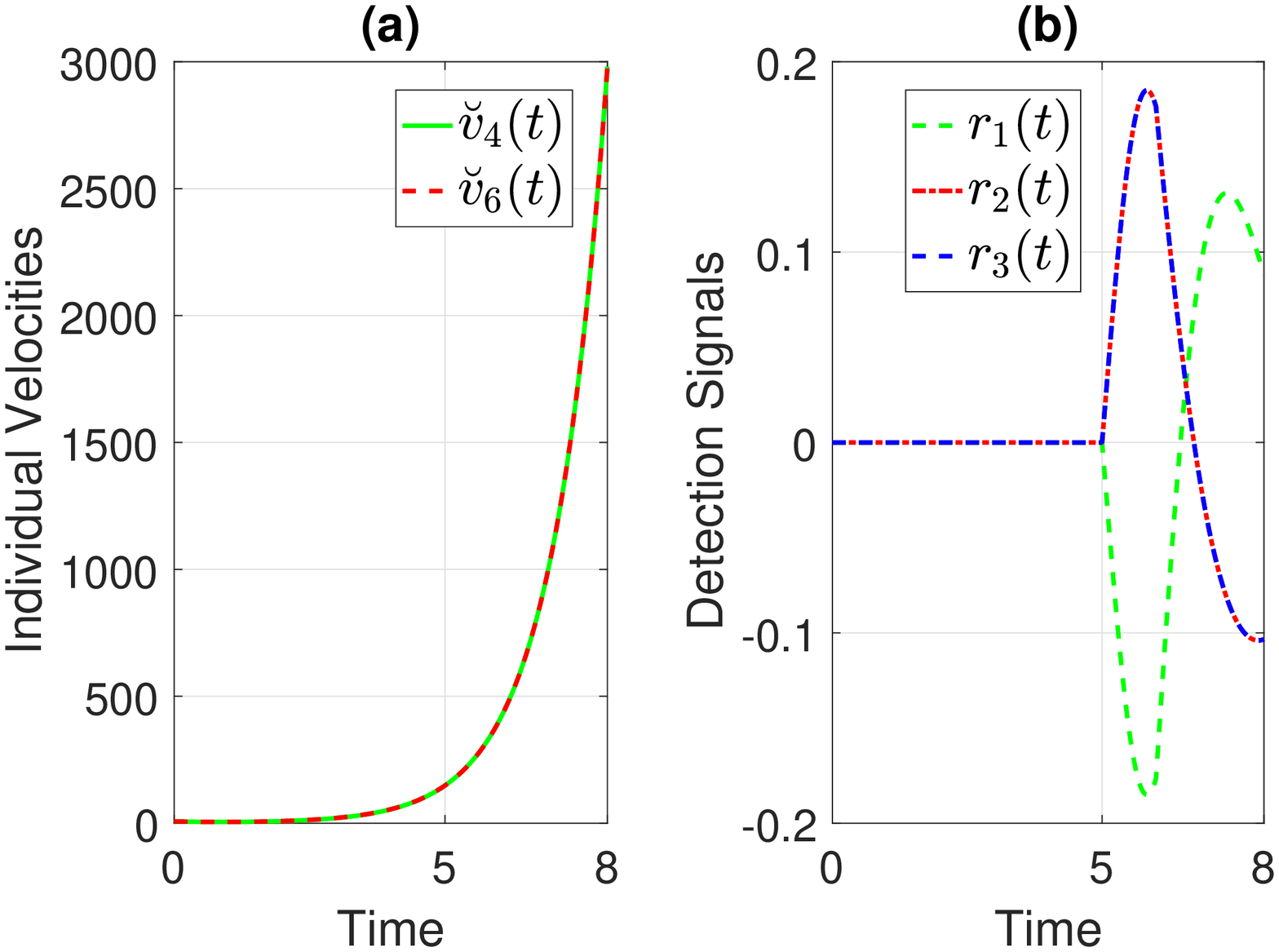}}
\caption{Trajectories of velocities (a)  and attack-detection signals (b).}
\label{fig:fs}
\end{figure}

The switching topologies in Figure~\ref{fig:tpc} (b) satisfy the defense strategy consisting of \eqref{eq:kkz0}, \eqref{eq:dfv} and \eqref{eq:kkz1x}. Therefore, with $y_{i}(t) = v_{i}(t)$, i.e., $c_{i1} = 0$, $i \in \mathbb{M} = \{1,2,3\}$, according to Theorem \ref{thm:thdb}, to detect the cooperative ZDA we can consider the periodic topology switching sequence in Figure~\ref{fig:tpc} (b): $\mathbf{L} = \left\{\underbrace{\sigma(t_0) = 5}_{\tau_{0} = 3}, \underbrace{\sigma(t_1) = 6}_{\tau_{1} = 1}\right\}.$ We assume that the attacker can modify any connection in the scope of attackable links. The trajectories of attack-detection signals in Figure~\ref{fig:fs} demonstrate that the observer \eqref{eq:fl} under Algorithm~1 succeeds in detecting the cooperative ZDA (nonzero detection signals).

\vspace{-0.50cm}
\subsection{Comparison with Existing Works}
The existing results on the detection of ZDA are summarized in Table~I. Since $|\mathbb{M}| = 1$ in Figure~\ref{fig:tpi} and $|\mathbb{K}| = 1$ for intermittent ZDA, $|\mathbb{M}| = 3$ in Figure~\ref{fig:tpc} and $|\mathbb{K}| = 6$ for cooperative ZDA, and the connectivity of all network topologies are the same as 1, which violate the conditions in Table~I.
Defense strategies that rely on only strategically changing system dynamics\cite{T12,ma1811}, while are effective against conventional ZDA and inspired us to analyze more sophisticated scenarios in this paper, implicitly assume that the attacker has no awareness of the aforementioned defense. Hence, the intermittent ZDA (when the system is unobservable) or cooperative ZDA (when the system is observable) cannot be detected by these methods. We also note that none of the prior work explicitly takes the issue of privacy/observability of initial/final states into account as we have pursued in this work.
\vspace{-0.20cm}
\begin{table}[http]
\caption{Conditions for Detection of ZDA}
\label{msft}
\centering
\scalebox{0.83}[0.83]{%
\begin{tabular}{  p{1.2cm}  |  p{6.0cm} |  p{2.0cm}}
\bottomrule
\cellcolor[gray]{0.9} \textbf{Reference} & \cellcolor[gray]{0.9} $ \cellcolor[gray]{0.9} \textbf{Conditions}$ &
\cellcolor[gray]{0.9} $\textbf{Dynamics}$\\  \hline
        \cite{F13}  & size of input-output linking is smaller than $|\mathbb{K}|$  & Continuous Time\\ \hline
        \cite{n11}   &connectivity is not smaller than $2|\mathbb{K}|$ + 1 &Discrete Time\\ \hline
        \cite{n12} & $|\mathbb{K}|$ is smaller than connectivity &Discrete Time \\ \hline
        \cite{weerakkody2017robust}  & the minimum vertex separator is larger than $|\mathbb{K}| + 1$  & Discrete Time\\ \hline
        \cite{chen2017protecting}  & single attack, i.e., $|\mathbb{K}|  = 1$  & Continuous Time\\
        \bottomrule
\end{tabular}
}
\end{table}
\vspace{-0.00cm}

\section{Conclusion}
In this paper, we have first introduced two ZDA variations for a scenario where the attacker is informed about the switching strategy of the defender: intermittent ZDA where the attacker pauses,  updates and resumes ZDA in conjunction with the knowledge of switching topologies, and cooperative ZDA where the attacker employs a stealthy topology attack to render the switching topology  defense ineffective. We have then studied conditions for a defender to detect these attacks, and subsequently based on these conditions, we have  proposed an   attack detection algorithm. The proposed defense strategy can detect both of the proposed ZDA variations, without requiring any knowledge of the set of misbehaving agents or the start, pause and resume times of the attack. Moreover, this strategy achieves asymptotic consensus and tracking in the absence of an attack without limiting the magnitudes of the coupling weights or the number of monitored agents.

Our analysis suggests an interesting  trade-off  among the switching cost, the duration of an undetected attack,  the convergence speed to consensus and tracking. Analyzing this fundamental trade-off through the lens of game theory and multi-objective optimization constitutes a part of our future research.

\appendices
\vspace{-0.20cm}
\section*{Appendix A: Auxiliary Lemmas}
In this section, we present auxiliary lemmas that are used in the proofs of the main results of this paper.

\begin{lem}~\cite{porfiri2008fast} Consider the switched systems:
\begin{align}
\dot x\left( t \right) = {\mathcal{A}_{\sigma \left( t \right)}}x\left( t \right)\nonumber
\end{align}
under periodic switching, i.e., $\sigma \left( t \right)$ $=$ $\sigma \left( {t + \tau} \right) \in \mathfrak{S}$.
If there exists a convex combination of some matrix measure that satisfies
\begin{align}
\sum\limits_{m = 0}^{l-1} {{\nu _m}\mu\left( {{\mathcal{A}_{m}}} \right)}  < 0,\label{eq:sdsbk1}
\end{align}
where ${\nu _m} = \frac{{{\tau _m}}}{{\sum\limits_{i = 0}^{l-1} {{\tau _i}} }}$; then the switched system system is uniformly asymptotically stable for every positive $\tau  = \sum\limits_{i = 0}^{l-1} {{\tau _i}}$.
\label{thm:pss}
\end{lem}

\begin{lem}~\cite{hom1991topics}
Consider the Vandermonde matrix:
\begin{align}
\mathcal{H} \triangleq \left[
    \begin{array}{c;{2pt/2pt}c;{2pt/2pt}c;{2pt/2pt}c}
        {1} & {1} & {\cdots} & {1}  \\ \hdashline[2pt/2pt]
        {a_1} & {a_2} & {\cdots} & {a_n}  \\ \hdashline[2pt/2pt]
        {a^2_1} & {a^2_2} & {\cdots} & {a^2_n}  \\ \hdashline[2pt/2pt]
        {\vdots} & {\vdots} & {\cdots} & {\vdots}  \\ \hdashline[2pt/2pt]
        {a^{n-1}_1} & {a^{n-1}_2} & {\cdots}  & {a^{n-1}_n}  \end{array}
\right] \in \mathbb{R}^{n \times n}. \nonumber
\end{align}
Its determinant is $\det \left( \mathcal{H} \right)  = {\left( { - 1} \right)^{\frac{{{n^2} - n}}{2}}}\prod\limits_{i < j} {\left( {{a_i} - {a_j}} \right)}.$
\label{thm:amm}
\end{lem}

\begin{lem}
Consider the matrix $Q_{r}$ that satisfies \eqref{eq:ufo4}. If ${\lambda _2}(\mathcal{L}_{r}) > 0$, then
\begin{align}
\ker \left( {\left[ {Q_r^\top} \right]_{{2:\left| \mathbb{V} \right|},:}} \right) = \left\{ {{\mathbf{1}_{\left| \mathbb{V} \right|}},{\mathbf{0}_{\left| \mathbb{V} \right|}}} \right\}. \label{eq:nddr}
\end{align}
\label{thm:ndd}
\end{lem}
\vspace{-0.40cm}
\begin{IEEEproof}
 The proof follows from a contradiction argument. We assume that \eqref{eq:nddr} does not hold, i.e., there exists a vector $\psi = [\varphi_{1}, \ldots,  \varphi_{\left| \mathbb{V} \right|}]^\top$ such that
\begin{equation}\label{eq:contradict}
\psi \notin  \emph{\emph{span}} \left\{ {{\mathbf{1}_{\left| \mathbb{V} \right|}},{\mathbf{0}_{\left| \mathbb{V} \right|}}} \right\},
\end{equation}
and
${\left[ {Q_r^\top} \right]_{{2:\left| \mathbb{V} \right|}, :}}\psi  = {\mathbf{0}_{\left| \mathbb{V} \right| - 1}}$. Then, it follows from \eqref{eq:ufo4} that
\begin{align}
{\mathcal{L}_r}\psi  = {Q_r}{\Lambda _r}Q_r^\top\psi  = {Q_r}{\mathbf{0}_{\left| \mathbb{V} \right|}} = {\mathbf{0}_{\left| \mathbb{V} \right|}}. \label{eq:addlm61}
\end{align}

From~\cite{ada11}, we know that an undirected graph is connected if and only if ${\lambda _2}(\mathcal{L}_{r}) > 0$, and further the null space of the Laplacian matrix $\mathcal{L}_r$ of a connected graph is spanned by the vector ${\mathbf{1}_{\left| \mathbb{V} \right|}}$. We obtain from \eqref{eq:addlm61} that ${\varphi_1} =  \ldots  = {\varphi_{\left| \mathbb{V} \right|}}$, which contradicts with \eqref{eq:contradict}. Thus, \eqref{eq:nddr} holds. This concludes the proof.
\end{IEEEproof}

\vspace{-0.10cm}
\section*{Appendix B: Proof of Proposition \ref{thm:zzaxx}}
Based on average variables $\bar x(t) \buildrel \Delta \over = \frac{1}{{\left| \mathbb{V} \right|}}\sum\limits_{i \in \mathbb{V}} {{x_i}\left( t \right)}$ and $\bar v(t) \buildrel \Delta \over = \frac{1}{{\left| \mathbb{V} \right|}}\sum\limits_{i \in \mathbb{V}} {{v_i}(t)}$, we define the following fluctuation terms:
\begin{subequations}
\begin{align}
{{{\tilde x}}_i}\left( t \right) &\triangleq {x_i}\left( t \right) - \bar{x}(t),\label{eq:flu1}\\
{{\tilde v}_i}\left( t \right) &\triangleq {v_i}\left( t \right) - \bar v(t),\label{eq:flu2}
\end{align}\label{eq:fluv}
\end{subequations}
\vspace{-0.20cm}
which implies that
\vspace{-0.00cm}
\begin{subequations}
\begin{align}
{\bf{1}}_{\left| \mathbb{V} \right|}^ \top \tilde x \left( t \right) &= 0, ~\text{for} ~ t \geq t_{0} \label{eq:usef1}\\
{\bf{1}}_{\left| \mathbb{V} \right|}^ \top \tilde v \left( t \right) &= 0, ~\text{for} ~ t \geq t_{0}. \label{eq:usef1no}
\end{align}\label{eq:j4}
\end{subequations}

\vspace{-0.40cm}
Considering \eqref{eq:oox2}, \eqref{eq:lci} and $a_{ij}^{\sigma(t)} = a_{ji}^{\sigma(t)}$, we have
\begin{align}
\dot{\bar v}(t) &= \frac{1}{{\left| \mathbb{V} \right|}}\sum\limits_{i \in \mathbb{V}} {{{\dot v}_i}(t)}  = \frac{1}{{\left| \mathbb{V} \right|}}\sum\limits_{i \in \mathbb{V}} {{u_i}(t)} \nonumber \\
&= \frac{1}{{\left| \mathbb{V} \right|}}\sum\limits_{i \in \mathbb{V}}( { - {v_i}(t) + \sum\limits_{j \in \mathbb{V}} {a_{ij}^{\sigma \left( t \right)}\left( {{x_j}\left( t \right) - {x_i}\left( t \right)} \right)} })\nonumber \\
& =  - \frac{1}{{\left| \mathbb{V} \right|}}\sum\limits_{i \in \mathbb{V}} {{v_i}(t)}  =  - \bar v(t), \nonumber
\end{align}which, in conjunction with \eqref{eq:flu2}, leads to
\begin{align}
&\dot{\tilde v}_i\left( t \right) \nonumber\\
& = {{\dot v}_i}\left( t \right) - \dot{\bar v}(t) = {u_i}\left( t \right) + \bar v(t) \nonumber\\
& =  - {v_i}(t) + \sum\limits_{j \in \mathbb{V}} {a_{ij}^{\sigma \left( t \right)}\left( {{x_j}\left( t \right) - {x_i}\left( t \right)} \right)}  + \bar v(t) \nonumber\\
& =  - ({v_i}(t) \!-\! \bar v(t)) \!+\! \sum\limits_{j \in \mathbb{V}}\!{a_{ij}^{\sigma(t)}( {( {{x_j}(t) \!-\! \bar x(t)}) \!-\! ( {{x_i}(t) \!-\! \bar x(t)})})} \nonumber\\
& =  - {{\tilde v}_i}(t) + \sum\limits_{j \in \mathbb{V}} {a_{ij}^{\sigma \left( t \right)}\left( {{{\tilde x}_j}\left( t \right) - {{\tilde x}_i}\left( t \right)} \right)} ,i \in \mathbb{V}.\label{real1}
\end{align}

The dynamics of the second-order multi-agent system \eqref{eq:ooiio} with control input \eqref{eq:lci} can now be expressed equivalently as
\begin{subequations}
\begin{align}
{\dot{\tilde{x}}}\left( t \right) &= {\tilde{v}}\left( t \right) \label{eq:foon1}\\
{{\dot{\tilde{v}}}}\left( t \right) &= - \tilde{v}\left( t \right) - {\mathcal{L}_{\sigma(t)}}\tilde{x}\left( t \right),\label{eq:foon2}
\end{align}\label{eq:foon}\end{subequations}
where \eqref{eq:foon2} considers its equivalent form \eqref{real1}.

Let us define $\hat x \triangleq Q_r^ \top \tilde x$  and $\hat v \triangleq Q_r^ \top \tilde v$. Noting \eqref{eq:ufo4}, the dynamics \eqref{eq:foon} can equivalently transform to
\begin{subequations}
\begin{align}
{\dot{\hat{x}}}\left( t \right) &= {\hat{v}}\left( t \right)\\
{{\dot{\hat{v}}}}\left( t \right) &= - \hat{v}\left( t \right) -{\Upsilon _{rs}}\hat{x}\left( t \right), r, s \in \mathbb{S}
\end{align}\label{eq:foonk}\end{subequations}
where $\Upsilon _{rs}$ is defined in \eqref{eq:sp1}.  We note that it follows from \eqref{eq:j4} and \eqref{eq:wp2} that ${{\hat x}_1}\left( t \right) = {{\hat v}_1}\left( t \right) = 0$, ${\left[ {{\Upsilon _{rs}}} \right]_{1, :}} = {\mathbf{0}^{\top}_{\left| \mathbb{V} \right|}}$ and ${\left[ {{\Upsilon _{rs}}} \right]_{:, 1}} = {\mathbf{0}_{\left| \mathbb{V} \right|}}$. Let us define $\theta \triangleq [\!\!\!
    \begin{array}{c;{1pt/1pt}c;{1pt/1pt}c;{1pt/1pt}c;{1pt/1pt}c;{1pt/1pt}c}
        \hat{x}_{2} \!&\! \ldots \!&\! \hat{x}_{\left| \mathbb{V} \right|} \!&\! \hat{v}_{2} \!&\! \ldots \!&\! \hat{v}_{\left| \mathbb{V} \right|}
    \end{array}
\!\!\!]^\top$. Thus, the system \eqref{eq:foonk} equivalently reduces to
\begin{align}
\dot \theta \left( t \right) = {\mathcal{A}_s}\theta \left( t \right),s \in \mathbb{S}\label{eq:xxcgf}
\end{align}
with ${\mathcal{A}_s}$ given in \eqref{eq:sp2}. Meanwhile, it is straightforward to verify that when $r = s$, $\mathcal{A}_{s}$ is Hurwitz. Therefore, the exists a $P >  0$ such that ${\mu_P}\left( \mathcal{A}_r \right)  <  0$. Through setting on the dwell time of the topology indexed by $r$, \eqref{eq:sdsbk1} can be satisfied. By Lemma~\ref{thm:pss},  the system \eqref{eq:xxcgf} is uniformly asymptotically stable, i.e., for any initial condition, $\mathop {\lim }\limits_{t \to \infty } \theta \left( t \right) = {\mathbf{0}_{2\left| \mathbb{V} \right| - 2}}$, which implies that $\mathop {\lim }\limits_{t \to \infty } {Q^\top}\tilde x\left( t \right) = \mathop {\lim }\limits_{t \to \infty } {Q^\top}\tilde v\left( t \right) = {\mathbf{0}_{\left| \mathbb{V} \right|}}$. Since $Q$ is full-rank, we have $\mathop {\lim }\limits_{t \to \infty } \tilde x\left( t \right) = \mathop {\lim }\limits_{t \to \infty } \tilde v\left( t \right) = {\mathbf{0}_{\left| \mathbb{V} \right|}}$. Then, \eqref{eq:fluv} implies that $\mathop {\lim }\limits_{t \to \infty } {{\tilde x}_i}\left( t \right) = \mathop {\lim }\limits_{t \to \infty } {{\tilde x}_j}\left( t \right)$ and $\mathop {\lim }\limits_{t \to \infty } {{\tilde v}_i}\left( t \right) = \mathop {\lim }\limits_{t \to \infty } {{\tilde v}_j}\left( t \right),\forall i \ne j \in \mathbb{V}$. Here, we conclude that the second-order consensus is achieved, and we define ${v^*} = \mathop {\lim }\limits_{t \to \infty } {{\tilde v}_i}\left( t \right),\forall i \in \mathbb{V}$. Then, substituting the second-order consensus into the system \eqref{eq:ooiio} with control input \eqref{eq:lci} yields the dynamics ${{\dot v}^*} =  - {v^*}$, which implies a common zero velocity at steady state.

\section*{Appendix C: Proof of Proposition \ref{thm:ccoo2}}
Let us first define:
\begin{align}
\mathbf y &\buildrel \Delta \over = \breve{y} - y.\label{eq:ed}
\end{align}
It is straightforward to obtain dynamics from \eqref{eq:sl2} and \eqref{eq:sl1} as
\begin{subequations}
\begin{align}
\dot{\mathbf z}\left( t \right) &= {A_{\sigma (t)}}\mathbf z\left( t \right) + \breve{g}\left( t \right) \\
\mathbf y\left( t \right) &= C\mathbf z\left( t \right) + D\breve{g}\left( t\right),
\end{align}\label{eq:psl2k2a}\end{subequations}
where ${\mathbf z}\left( t \right)$ is defined in \eqref{eq:ed0}.

\subsubsection{\textbf{Proof of \eqref{eq:rs2as}}} Since $\left[ {{{\xi_k}},{\zeta_{k}}} \right) \subseteq \left[{{t_k},{t_{k + 1}}}\right)$, $\sigma(t) = r$ for $t \in \left[ {{{\xi_k}},{\zeta_{k}}} \right)$.
We denote ${\Xi}\left( s \right) \triangleq \mathfrak{L}\left\{ \mathbf z\left( t \right) \right\}$, where $\mathfrak{L}(\cdot)$ stands for the Laplace transform operator. It follows from the attack signal \eqref{eq:ads} that $\mathfrak{L}\left\{ {\breve{g}(t)} \right\} = {(e^{ - {{\xi_k}}s} - e^{ {-\zeta_{k}}s})}\frac{{\breve{g}\left( {{{\xi_k}}} \right)}}{{s - {\eta_{r}}}}$, $t \in \left[ {{{\xi_k}},{\zeta_{k}}} \right)$. Without loss of generality, we let $\sigma(t) = r$ for $t \in [t_{k}, t_{k+1})$.  Then, the Laplace transform of the dynamics in \eqref{eq:psl2k2a} is obtained as
\begin{align}
&{(e^{ - {{\xi_k}}s} - e^{ {-\zeta_{k}}s})}\!\left( {s\Xi \left( s \right) - \mathbf z\!\left( {{{\xi _k}}} \right)} \right) \nonumber\\
&= {(e^{ - {{\xi_k}}s} - e^{ {-\zeta_{k}}s})}{A_{r}}\Xi \left( s \right) + {(e^{ - {{\xi_k}}s} - e^{ {-\zeta_{k}}s})}\frac{{\breve{g}\!\left( {{{\xi_k}}} \right)}}{{s - {\eta_{r}}}},\nonumber
\end{align}
which is equivalent to
\begin{align}
\!\!{(e^{ - {{\xi_k}}s} \!-\! e^{ {-\zeta_{k}}s})}\Xi \!\left(\! s\! \right) \!=\! \frac{{(e^{ - {{\xi_k}}s} \!-\! e^{ {-\zeta_{k}}s})}}{{s{{\bf{1}}_{2\left| \mathbb{V} \right| \times 2\left| \mathbb{V} \right|}} \!-\! {A_r}}}\!\!\left(\!\! {\mathbf z\left( {{{\xi _k}}} \right) \!+\! \frac{{\breve{g}\!\left( {{{\xi _k}}} \right)}}{{s \!-\! {\eta_{r}}}}} \!\!\right)\!\!.\label{eq:eqz3}
\end{align}

Expanding \eqref{eq:ap2} out yields
\begin{align}
C\mathbf z\left( {{{\xi_k}}} \right) + {D}\breve{g}\left( {{\xi_{k}}} \right) &= {\mathbf{0}_{\left| \mathbb{M} \right|}},\label{eq:eqz4a}\\
{\eta_{r}}\mathbf z\left( {{{\xi_k}}} \right) - {A_r}\mathbf z\left( {{\xi_{k}}} \right) &= \breve{g}\left( {{\xi_{k}}} \right),r \in \mathbb{T}.\label{eq:eqz4b}
\end{align}Substituting \eqref{eq:eqz4b} into \eqref{eq:eqz3} yields $(e^{ - {{\xi_k}}s} - e^{ {-\zeta_{k}}s})\Xi \left( s \right) = \frac{(e^{ - {{\xi_k}}s} - e^{ {-\zeta_{k}}s})}{{s - {\eta_{r}}}}\mathbf z\left( {{{\xi _k}}} \right)$, and
the inverse Laplace transform of it gives \eqref{eq:rs2as}.

\subsubsection{\textbf{Proof of \eqref{eq:cad}}} It follows from \eqref{eq:rs2as} and \eqref{eq:psl2k2a} that
\begin{align}
\mathbf y\!\left( t \right) \!=\! {e^{\eta_{r}\left( {t - {{\xi_k}}} \right)}}\left( {C\mathbf{z}\!\left( {{{\xi_k}}} \right) \!+\! {D}\breve{g}({{\xi _k}})} \right),t \!\in\! \left[ {{{\xi _k}},{\zeta_k}} \right), k \!\in\! \mathbb{N}_{0}\label{eq:ppk}
\end{align}
which combined with \eqref{eq:eqz4a} results in $\mathrm y\left( t \right) = \mathbf{0}_{\left| \mathbb{M} \right|}$, or equivalently, $\breve{y}\left( t \right) = {y}\left( t \right)$, for any $t \in \left[ {{{\xi _k}},{{\zeta_k}}} \right)$.

We next prove \eqref{eq:rs1} over non-attack interval of ZDA $\left[ {\zeta_{k}, {\xi_{k + 1}}} \right)$. From \eqref{eq:ads} and \eqref{eq:ao2}, the dynamics \eqref{eq:psl2k2a} over such non-attack intervals of ZDA (subject to the monitored output attack as \eqref{eq:ao2}) is described by
\begin{subequations}
\begin{align}
\dot{\mathbf z}\left( t \right) &= {A_{\sigma (t)}}\mathbf z\left( t \right)\label{eq:nap21a}\\
\mathbf y\left( t \right) &= C\mathbf z\left( t \right) + D\sum\limits_{m = 0}^{k} {\breve{g}}\left( \zeta^{-}_{m} \right),t \in \left[ {\zeta_{k}, {\xi_{k + 1}}} \right).\label{eq:nap21b}
\end{align}\label{eq:nap21}\end{subequations}
It follows from \eqref{eq:rs2as} and \eqref{eq:nap21a} that
\begin{align}
&\mathbf z(t)\label{eq:nap3}\\
& = \! \left\{ \begin{array}{l}
\hspace{-0.22cm}{e^{{A_{\sigma \left( {{t_k}} \right)}}\left( {t - {t_k}} \right)}}\mathbf z\left( {{t_k}} \right),\hspace{3.45cm}t \!\in\!\! \left[ {{t_k},{{\xi_k}}} \right)\\
\hspace{-0.22cm}{e^{{\mathbf{1}_{2|\mathbb{V}| \times 2|\mathbb{V}|}}\eta_{r}\left( {t - {{\xi _k}}} \right) + {A_{\sigma \left( {{t_k}} \right)}}\left( {{{\xi _k}} - {t_k}} \right)}}\mathbf z\left( {{t_k}} \right),\hspace{0.73cm}t \!\in\!\! \left[ {{{\xi_k}},{\zeta _k}} \right)\\
\hspace{-0.22cm}{e^{{A_{\sigma \!\left(\! {{t_k}} \!\right)}}\!\left(\! {t - {t_k} - \left( {{\zeta _k} - {{\xi _k}}} \right)} \!\right) + {\mathbf{1}_{2|\mathbb{V}| \times 2|\mathbb{V}|}}\eta_{r}\left( {{\zeta _k} - {{\xi _k}}} \right)}}\mathbf z\!\left(\! {{t_k}} \!\right)\!, t \!\in\!\! \left[ {{\zeta _k},{t_{k + 1}}} \!\right)\!.
\end{array} \right. \nonumber
\end{align}

We conclude from \eqref{eq:ed} that \eqref{eq:cad} is equivalent to
\begin{align}
\mathbf{y}\left( t \right) \equiv \mathbf{0}_{\left| \mathbb{M} \right|} ~\text{on}~ \left[ {{{t_0}},{{t_{k+1}}}} \right). \label{vro1}
\end{align}

For $D = {\mathbf{0}_{\left| \mathbb{M} \right| \times 2\left| \mathbb{V} \right|}}$, we note that \eqref{vro1} implies that the system \eqref{eq:nap21} is unobservable for any $t \in \left[ {{t_0},{t_{k + 1}}} \!\right)$, $k \in \mathbb{N}_{0}$. It is immediate that
\begin{align}
\mathbf{z}(t_{k}) \in  \ker ({\mathcal{O}}_{k} ) = \widehat{\mathbf{N}}^{k}_{k}, ~k \in \mathbb{N}_{0}. \label{vro2}
\end{align}
We next show that $\mathbf{z}(t_{q-1}) \in  \widehat{\mathbf{N}}^{k}_{q-1}$ for $0 \leq q-1 \leq k$, through inductive argument. Let us suppose $\mathbf{z}(t_{q}) \in  \widehat{\mathbf{N}}^{k}_{q}$. We obtain from \eqref{eq:nap3} that $\mathbf{z}( {{t_q}})$ $=$ $\mathbf{z}(t^{-}_q)$ $=$ ${e^{{\eta _{\sigma (t_{q - 1})}}( {{\zeta _{q - 1}} - {\xi _{q - 1}}})}}{e^{{A_{\sigma ({{t_{q - 1}}})}}( {{\tau _{q - 1}} - ( {{\zeta _{q - 1}} - {\xi _{q - 1}}})})}}\mathbf{z}( {{t_{q - 1}}})$, which, in conjunction with the fact of $e^{{\eta _{\sigma (t_{q - 1})}}( {{\zeta _{q - 1}} - {\xi _{q - 1}}})} \neq 0$, leads to
$\mathbf{z}(t_{q-1})$ $\in$  ${e^{{-A_{\sigma( {{t_{q - 1}}})}}( {{\tau _{q - 1}} - ( {{\zeta _{q - 1}} - {\xi _{q - 1}}})})}}\widehat{\mathbf{N}}^{k}_{q}$. Moreover, we note that \eqref{vro2} implies that $\mathbf{z}(t_{q-1}) \in  \ker ({\mathcal{O}}_{q-1} )$. Therefore,
\begin{align}
\!\!\!\!\mathbf{z}(t_{q-1}) \!\in\!  {e^{{-A_{\sigma(t_{q - 1})}}({{\tau _{q - 1}} - ( {{\zeta _{q - 1}} - {\xi _{q - 1}}})})}}\widehat{\mathbf{N}}^{k}_{q} \!\cap\! \ker ({\mathcal{O}}_{q-1}), \label{vro3}
\end{align}
where the right-hand expression is, in fact, the computation of $\widehat {\bf{N}}_{q - 1}^k$, i.e., the unobservable space given by \eqref{eq:cm2}. Let $q = 1$, we have $\mathbf{z}(t_{0}) \in \widehat {\bf{N}}_{0}^k$. Then, following the same steps in the proof of necessary condition in Theorem 1 of~\cite{tanwani2013observability}, we conclude that \eqref{eq:cad} holds if and only if there exists a non-zero vector $\mathbf{z}\left( {{t_0}} \right)$ such that
\begin{align}
\mathbf{z}\left( {{t_0}} \right) \in  \widehat{\mathbf{N}}^{k}_{0}. \label{eq:ao1c}
\end{align}

For $D \neq {\mathbf{0}_{\left| \mathbb{M} \right| \times 2\left| \mathbb{V} \right|}}$, it follows from \eqref{eq:eqz4a} and \eqref{eq:nap21b} that $\mathbf y\left( {{\zeta _k}} \right) = \mathbf y\left( {\zeta _k^ - } \right) = {\mathbf{0}_{\left| \mathbb{M} \right|}}$. Therefore, in this scenario, \eqref{vro1} holds only when $\dot{\mathbf{y}}\left( t \right) \equiv \mathbf{0}_{\left| \mathbb{M} \right|} ~\text{on}~ \left[ {{{t_0}},{{t_{k+1}}}} \right)$. Updating the observability matrix ${\mathcal{O}}_{q}$ in \eqref{eq:om1} by ${\widetilde{\mathcal{O}}}_{q}$ in \eqref{eq:om1poa} and following the same steps to derive \eqref{eq:ao1c},  we conclude that \eqref{eq:cad} holds if and only if
\vspace{-0.1cm}
\begin{align}
\mathbf{z}\left( {{t_0}} \right) \in  \widetilde{\mathbf{N}}^{k}_{0}, \label{eq:ao2c}
\end{align}
where $\widetilde{\mathbf{N}}^{k}_{0}$ is recursively computed by~(\ref{eq:cm3}) and~(\ref{eq:cm4}).

In addition to \eqref{eq:ao1c} and \eqref{eq:ao2c}, we conclude that if \eqref{eq:ap1} and \eqref{eq:eqz4a} hold, regardless of $D\sum\limits_{m = 0}^{k} {\breve{g}}\left( \zeta^{-}_{m} \right)$ $\neq {\mathbf{0}_{\left| \mathbb{M} \right|}}$ or $= {\mathbf{0}_{\left| \mathbb{M} \right|}}$, \eqref{eq:cad} always holds.

\section*{Appendix D: Proof of Proposition \ref{thm:nnq}}
Let us define $\tilde{e} \triangleq \left[\!\!\!
    \begin{array}{c;{1pt/1pt}c}
        {\tilde{e}_x^\top} & {\tilde{e}_v^\top}
    \end{array}\!\!\!
\right]^\top$ $\triangleq$ $\widehat{z} - z$. Without loss of generality, we let $\sigma(t_{k+1}) = s$. Noticing \eqref{eq:kkc}, we obtain from the dynamics \eqref{eq:sasb} and \eqref{eq:s2} that
\begin{subequations}
\begin{align}
\!\!\! \dot{\tilde{e}}(t) &= {{\widehat A}_{s}}\tilde{e}(t) \!+\! ({{{\widehat A}_{s}} \!-\! {A_{s}}} )z( t),t \!\in\! \left[ {{t_{k + 1}},{t_{k + 2}}} \right) \label{eq:po1}\\
\!\!\! \widehat{{y}}( t) \!-\! y(t) &= C\tilde{e}(t), \label{eq:po2}\\
\!\!\! \tilde{e}(t_{k+1}) &= \mathbf{0}_{\left| \mathbb{M} \right|}, \label{eq:po3x}
\end{align}\label{eq:pox}\end{subequations}
from which we have
\begin{align}
\widehat{{y}}(t) \!-\! y(t) \!=\! C{e^{{\widehat{A}_s}( {t - {t_{k + 1}}})}}\!\!\!\int_{{t_{k + 1}}}^t \!\!\!\!\!{{e^{ - \widehat{A}_{s}( {\tau  - {t_{k + 1}}})}}({(\!{{{\widehat A}_s} \!-\! {A_s}} \!)z( \tau  )} )\mathrm{d}\tau, }\nonumber
\end{align}
and the corresponding derivatives
\begin{align}
&\widehat{{y}}^{(d)}(t) - {y^{(d)}}(t)\nonumber\\
& = C\widehat{A}_s^d{e^{{\widehat{A}_s}( {t - {t_{k + 1}}})}}\!\!\!\int_{{t_{k + 1}}}^t \!\!\!\!\!{{e^{ - \widehat{A}_{s}( {\tau  - {t_{k + 1}}})}}{( {{{\widehat A}_s} \!-\! {A_s}} )z( \tau )}\mathrm{d}\tau } \nonumber\\
&\hspace{2.4cm} + \sum\limits_{l = 0}^{d - 1} {C\widehat{A}_s^l( {( {{{\widehat A}_s} - {A_s}}){z^{( {d - 1 - l} )}}( t)})}. \label{eq:po3x}
\end{align}

We note that under corrupted topology, the stealthy property $\widehat{{y}}\left( t \right) - y\left( t \right) = {\mathbf{0}_{\left| \mathbb{M} \right|}}$ for any $t \in \left[ {{t_{k + 1}},{t_{k + 2}}} \right)$ is equivalent to $\widehat{{y}}^{\left( d \right)}\left( t_{k+1} \right) - {y^{\left( d \right)}}\left( t_{k+1} \right) = \mathbf{0}_{\left| \mathbb{M} \right|}$ for $\forall d \in {\mathbb{N}_0}$, which is further equivalent to \eqref{eq:xxz} by considering the solution \eqref{eq:po3x}.

\section*{Appendix E: Proof of Theorem \ref{thm:thd}}
Without loss of generality, we let $\sigma(\zeta _k) = r \in \mathbb{T}$, and $\zeta _k < t_{k+1}$, $k \in \mathbb{N}$, i.e., attacker ``pauses" ZDA at $\zeta_{k}$. We now prove this theorem via a contradiction.  We assume that the attack is not detectable in $[\zeta^-_{k}, \xi_{k+1})$,  which is equivalent to
\begin{align}
{{\mathbf{y}}}\left( t \right) = {\mathbf{0}_{\left| \mathbb{M} \right|}} ~\text{for any}~ t \in [{{\zeta^-_k}},{\xi_{k + 1}}), \label{eq:ed1}
\end{align}where ${{\mathbf{y}}}\left( t \right)$ is defined in \eqref{eq:ed}.

Considering the fact that given a differentiable function $f(t)$, $f(t)=0$ for any $t \in [a,b]$, if and only if $f(a)=0$ and $f^{(d)}(a)=0,~\forall d\in \mathbb{N}$. We conclude from \eqref{eq:nap21} that \eqref{eq:ed1} at time $\zeta_{k}$ is equivalent to
\begin{align}
{{\bf{y}}^{\left( d \right)}}\!\!\left( {{\zeta _k}} \right) \!=\! \left\{ \begin{array}{l}
\hspace{-0.2cm}C{\bf{z}}\left( {{\zeta _k}} \right) + D\!\!\sum\limits_{m = 0}^{k} {\breve{g}}\left( \zeta^{-}_{m} \right) = {\mathbf{0}_{\left| \mathbb M \right|}},\hspace{0.25cm}d = 0\\
\hspace{-0.2cm}C{A_{r}^d}{\bf{z}}\left( {{\zeta _k}} \right) = {\mathbf{0}_{\left| \mathbb M \right|}},\hspace{1.90cm}\forall d \in \mathbb{N}.
\end{array} \right.\label{eq:pk1}
\end{align}

With the definitions of  $A_{r}$, $C$, $D$ and ${\bf{z}}\left( \cdot \right)$ in \eqref{eq:nm0}, \eqref{eq:nmok}, \eqref{eq:nmokbb} and \eqref{eq:ed0}, the relation \eqref{eq:pk1} can be further rewritten under different forms of observation as follows:
\begin{itemize}
  \item Full Observation of Velocity, i.e., $c_{i1} = 0$, $\forall i \in \mathbb{M}$,
 \begin{subequations}
\begin{align}
{C_2}\mathbf{v}\left( {{\zeta _k}} \right) + {D}\!\!\sum\limits_{m = 0}^{k} {\breve{g}}\left( \zeta^{-}_{m} \right) &= {\mathbf{0}_{\left| \mathbb{M} \right|}} \label{eq:fv1}\\
{C_2}\mathbf{v}\left( {{\zeta _k}} \right) + {C_2}{\mathcal{L}_r}\mathbf{x}\left( {{\zeta _k}} \right) &= {\mathbf{0}_{\left| \mathbb{M} \right|}} \label{eq:fv2}\\
{C_2}\mathcal{L}_r^e\mathbf{v}\left( {{\zeta _k}} \right) &= {\mathbf{0}_{\left| \mathbb{M} \right|}},\forall e \in \mathbb{N} \label{eq:fv3}\\
{C_2}\mathcal{L}_r^d\mathbf{x}\left( {{\zeta _k}} \right) &= {\mathbf{0}_{\left| \mathbb{M} \right|}},\forall d \in {\mathbb{N}_{ \ge 2}} \label{eq:fv4}
\end{align}\label{eq:fv}\end{subequations}
\vspace{-0.50cm}
  \item Full Observation of Position, i.e., $c_{i2} = 0$, $\forall i \in \mathbb{M}$,
  \begin{subequations}
\begin{align}
{C_1}\mathbf{x}\left( {{\zeta _k}} \right) + D\!\!\sum\limits_{m = 0}^{k} {\breve{g}}\left( \zeta^{-}_{m} \right) &= {\mathbf{0}_{\left| \mathbb{M} \right|}}   \label{eq:fp1}\\
{C_1}\mathcal{L}_r^e\mathbf{x}\left( {{\zeta _k}} \right) &= {\mathbf{0}_{\left| \mathbb{M} \right|}},\forall e \in \mathbb{N}\label{eq:fp2}\\
{C_1}\mathcal{L}_r^d\mathbf{v}\left( {{\zeta _k}} \right) &= {\mathbf{0}_{\left| \mathbb{M} \right|}},\forall d \in {\mathbb{N}_0}\label{eq:fp3}
\end{align}\label{eq:fp}\end{subequations}
\vspace{-0.50cm}
\item Partial Observation, i.e., $c_{i1} \neq 0$ and $c_{i2} \neq 0$, $\forall i \in \mathbb{M}$,
  \begin{subequations}
\begin{align}
\!\!\!\!\!\!{C_1}\mathbf{x}\left( {{\zeta _k}} \right) \!+\! {C_2}\mathbf{v}\left( {{\zeta _k}} \right) \!+\! D\!\!\sum\limits_{m = 0}^{k} {\breve{g}}\left( \zeta^{-}_{m} \right) &\!=\! {\mathbf{0}_{\left| \mathbb{M} \right|}},\label{eq:po1}\\
\!\!\!\!\!\!\!\!\!\!{C_1}\mathcal{L}_r^e\mathbf{x}\left( {{\zeta _k}} \right) + {C_2}\mathcal{L}_r^e\mathbf{v}\left( {{\zeta _k}} \right) &\!=\! {\mathbf{0}_{\left| \mathbb{M} \right|}},\forall e \!\in\! \mathbb{N}\label{eq:po2}\\
\!\!\!\!\!\!\!\!\!\!\!\!\!\!\!\left( {{C_1} \!-\! {C_2}} \right)\!\mathcal{L}_r^d\mathbf{v}\!\left( {{\zeta _k}} \right) \!-\! {C_2}\mathcal{L}_r^{d + 1}\mathbf{x}\!\left( {{\zeta _k}} \right) &\!=\! {\mathbf{0}_{\left| \mathbb{M} \right|}},\forall d \!\in\! {\mathbb{N}_0}. \label{eq:po3}
\end{align}\label{eq:po}\end{subequations}
\end{itemize}

\vspace{-0.50cm}
Considering the definition of the vector $\mathbf{z}(t)$ in \eqref{eq:ed0}, and its continuity with respect to time, i.e., $\mathbf{z}\left( {{{\zeta^{-}_k}}} \right) = \mathbf{z}\left( {{{\zeta_k}}} \right)$, it follows from \eqref{eq:rs2as} and \eqref{eq:ads} that at time ${\zeta^{-}_k}$,
\vspace{-0.20cm}
\begin{align}
\left[
    \begin{array}{c}
        \mathbf{z}\left( {{{\zeta_k}}} \right)\\ \hdashline[2pt/2pt]
        -\breve{g}\left( {{{\zeta^{-}_k}}} \right)
    \end{array}
\right] = {e^{{\eta_{r}}\left( {{{\zeta^{-}_k}} - {{\xi_k}}} \right)}}\left[
    \begin{array}{c}
        \mathbf{z}\left( {{{\xi _k}}} \right)  \\ \hdashline[2pt/2pt]
        -\breve{g}\left( {{{\xi_k}}} \right)
    \end{array}
\right], \label{eq:pk22a}
\end{align}
which, in conjunction with the fact of ${e^{{\eta_{r}}\left( {{{\zeta^{-}_k}} - {{\xi _k}}} \right)}} \neq 0$ and the condition \eqref{eq:ap2}, results in
\begin{align}
\left[
    \begin{array}{c}
        \mathbf{z}\left( {{{\zeta_k}}} \right)\\ \hdashline[2pt/2pt]
        -\breve g\left( {{{\zeta^{-}_k}}} \right)
    \end{array}
\right] \in \ker \left( {{\mathcal{P}_k}} \right). \label{eq:pk2}
\end{align}

With variables $\breve{g}(\zeta^{-}_k)$, $\bar{g}(\zeta^{-}_k)$, $\mathbf{z}\left( {{{\zeta_k}}} \right)$, $A_{r}$ and $\mathcal{P}_k$ defined in \eqref{eq:nm2}, \eqref{eq:u1}, \eqref{eq:ed0}, \eqref{eq:nm0} and \eqref{eq:pal21}, respectively, expanding \eqref{eq:pk2} yields
\begin{align}
\eta_{r}\mathbf{x}\left( \zeta_k \right) - \mathbf{v}\left( \zeta_k \right) &= \mathbf{0}_{\left| \mathbb{V} \right|},\label{eq:ss1}\\
-{\bar{g}}\left( \zeta^{-}_k \right) + \mathbf{v}(\zeta_k) + {\mathcal{L}_r}\mathbf{x}\left( \zeta_k \right) + \eta_{r} \mathbf{v}(\zeta_k) &= \mathbf{0}_{\left| \mathbb{V} \right|}.\label{eq:ss2}
\end{align}

Before proceeding the rest of proof, we define the variables:
\begin{subequations}
\begin{align}
H_{i} &\triangleq [{{\cal U}_{ri}}\mathbf{x}\left( {{\zeta _k}}\right)]_{2:|\mathbb{V}|},  \label{eq:pth29}\\
{\mathcal{D}_r} &\triangleq \emph{\emph{diag}}\left\{ {{\lambda^2_2}\left( {{\mathcal{L}_r}} \right), \ldots ,{\lambda^2_{\left| \mathbb{V} \right|}}\left( {{\mathcal{L}_r}} \right)} \right\},\label{eq:pth28cc}\\
\widetilde{\mathcal{H}}_{r} &\triangleq \left[
    \begin{array}{c;{2pt/2pt}c;{2pt/2pt}c}
        {{{{\lambda^2_2}({\mathcal{L}_r})}}} & {\cdots} & {{{ {\lambda^2_{\left| \mathbb{V} \right|}}({\mathcal{L}_\mathrm{r}})}}}  \\ \hdashline[2pt/2pt]
        {{{{\lambda^3_2}({\mathcal{L}_r})}}} & {\cdots} & {{{ {\lambda^3_{\left| \mathbb{V} \right|}}({\mathcal{L}_r})}}}  \\ \hdashline[2pt/2pt]
        {\vdots} & {\cdots} & {\vdots}  \\ \hdashline[2pt/2pt]
        {{ {{\lambda^{\left| \mathbb{V} \right|}_2}({\mathcal{L}_r})}}} & {\cdots}  & {{{{\lambda^{\left| \mathbb{V} \right|}_{\left| \mathbb{V} \right|}}({\mathcal{L}_r})}}}   \end{array}
\right],\label{eq:pth27}\\
\mathcal{H}_{r} &\triangleq \left[
    \begin{array}{c;{2pt/2pt}c;{2pt/2pt}c}
        {{{1}}} & {\cdots} & {{{ {1}}}}  \\ \hdashline[2pt/2pt]
        {{{{\lambda_2}({\mathcal{L}_r})}}} & {\cdots} & {{{ {\lambda_{\left| \mathbb{V} \right|}}({\mathcal{L}_r})}}}  \\ \hdashline[2pt/2pt]
        {\vdots} & {\cdots} & {\vdots}  \\ \hdashline[2pt/2pt]
        {{ {{\lambda^{\left| \mathbb{V} \right|-2}_2}({\mathcal{L}_r})}}} & {\cdots}  & {{{{\lambda^{\left| \mathbb{V} \right|-2}_{\left| \mathbb{V} \right|}}({\mathcal{L}_r})}}}   \end{array}
\right],\label{eq:pth26}
\end{align}
\end{subequations}where ${\mathcal{U}_{ri}}$ is given in \eqref{eq:da}.

\vspace{-0.40cm}
\subsection{Under Full Observation of Position or Velocity}
\vspace{-0.10cm}
Let us start with full observation of velocity. It follows from \eqref{eq:ufo4} that ${\cal L}_r^d = {Q_r}\Lambda _r^dQ_r^ \top$ with $\Lambda _r$ given in \eqref{eq:ufo4}. Thus, \eqref{eq:fv4} is equivalent to  ${C_2}{Q_r}\Lambda _r^dQ_r^ \top \mathbf{x}\left( {{\zeta _k}} \right) = {{\bf{0}}_{\left| \mathbb M \right|}},\forall d \in {\mathbb N_{ \ge 2}}$, which is further equivalent to
\begin{align}
\sum\limits_{l = 1}^{\left| \mathbb{V} \right|} {{{{{\lambda^{d}_l}({\mathcal{L}_r})}}}{\left[ {{Q_r}} \right]_{i,l}}\left[Q_r^\top\right]_{l,:} \mathbf{x}\left({\zeta_{k}} \right)}  = 0,\forall d \in {\mathbb{N}}, \forall i \in \mathbb{M} \label{eq:xxc}
\end{align}
with the consideration of the matrix $C_2$ defined in \eqref{eq:u1b} with $c_{i2} \neq 0, \forall i \in \mathbb{M}$. Further, recalling $\widetilde{\mathcal{H}}_r$, $H_{i}$ and ${\mathcal{U}_{ri}}$ from \eqref{eq:pth27}, \eqref{eq:pth29} and \eqref{eq:da}, from \eqref{eq:xxc} we have
\begin{align}
{\widetilde{\mathcal{H}}_r}H_{i} = {\mathbf{0}_{\left| \mathbb{V} \right|-1}}, \forall i \in \mathbb{M}.\label{eq:ptha1}
\end{align}

It can be verified from \eqref{eq:pth28cc}--\eqref{eq:pth26} that ${\widetilde{\mathcal{H}}_r} = {\mathcal{H}_r}{\mathcal{D}_r}$, from which we have $\det ( {{\widetilde{\mathcal{H}}_r}} )$ $=$ $\det ( {{\mathcal{H}_r}})\det( {{\mathcal{D}_r}} )$. The matrix defined in \eqref{eq:pth28cc} shows if $\mathcal{L}_{r}$ has distinct eigenvalues, $\mathcal{D}_r$ is full-rank. In addition, by Lemma~\ref{thm:amm}, the Vandermonde matrix $\mathcal{H}_{\mathrm{r}}$ is full-rank; thus, ${\widetilde{\mathcal{H}}_r}$ is full-rank. Therefore, the solution of \eqref{eq:ptha1} is
\begin{align}
H_{i} = {\mathbf{0}_{\left| \mathbb{V} \right|-1}}, \forall i \in \mathbb{M}.\label{eq:ptha2}
\end{align}

With the definitions in \eqref{eq:da} and \eqref{eq:pth29}, the equation \eqref{eq:ptha2} indicates that for $\forall i \in \mathbb{M}$,
\begin{align}
\!\!\text{diag}\!\left\{\! {{{\left[ {{Q_r}} \right]}_{i,2}}, \ldots ,{{\left[ {{Q_r}} \right]}_{i,\left| \mathbb{V} \right|}}} \!\right\}\!\!{\left[ {Q_r^\top } \right]_{2:\left| \mathbb{V} \right|, :}}\mathbf{x}\left({\zeta_{k}} \right) \!=\! {\mathbf{0}_{\left| \mathbb{V} \right| - 1}}.\label{eq:ko}
\end{align}

We note that \eqref{eq:da}, \eqref{eq:dahh} and \eqref{eq:kkz1} imply that $\exists i \in\mathbb{M}:$ $\text{diag}\left\{ {{{\left[ {{Q_r}} \right]}_{i,2}}, \ldots ,{{\left[ {{Q_r}} \right]}_{i,\left| \mathbb{V} \right|}}} \right\}$ is full-rank. Thus, from \eqref{eq:ko} we have ${\left[ {Q_r^\top } \right]_{{2:\left| \mathbb{V} \right|}, :}}\mathbf{x}\left({\zeta_{k}} \right) = {\mathbf{0}_{\left| \mathbb{V} \right| - 1}}$. By Lemma~\ref{thm:ndd}, the solution of \eqref{eq:ko} is
\begin{align}
{\mathbf{x}_1}\left( {{{\zeta_k}}} \right) =  \ldots  = {\mathbf{x}_{\left| \mathbb{V} \right|}}\left( {{{\zeta_k}}} \right). \label{eq:pthb1}
\end{align}
Considering \eqref{eq:fv3}, using the same method to derive \eqref{eq:pthb1}, we obtain \begin{align}
{\mathbf{v}_1}\left( {{{\zeta_k}}} \right) =  \ldots  = {\mathbf{v}_{\left| \mathbb{V} \right|}}\left( {{{\zeta_k}}} \right). \label{eq:aq1}
\end{align}

Substituting \eqref{eq:pthb1} into \eqref{eq:fv2}  yields ${C_2}\mathbf{v}\left( {{\zeta _k}} \right) = {\mathbf{0}_{\left| \mathbb{M} \right|}}$, which together with \eqref{eq:aq1}  results in
\begin{align}
{\mathbf{v}_1}\left( {{{\zeta_k}}} \right) =  \ldots  = {\mathbf{v}_{\left| \mathbb{V} \right|}}\left( {{{\zeta_k}}} \right) = 0. \label{eq:t2}
\end{align}

For the full observation of position, using nearly the same analysis method employed above, we obtain the same results as \eqref{eq:pthb1} and \eqref{eq:t2}.

Substituting \eqref{eq:pthb1} and \eqref{eq:t2}
into \eqref{eq:ss2} yields ${\bar{g}}\left( \zeta^{-}_k \right) = \mathbf{0}_{\left| \mathbb{V} \right|}$, and consequently, ${\breve{g}}\left( \zeta^{-}_k \right) = \mathbf{0}_{2\left| \mathbb{V} \right|}$. This means that there is no ZDA on the system at $\zeta^{-}_k$, which contradicts the assumption that the attack is applied until $\zeta_k$.
Therefore, we conclude that under the full observation of position or velocity, the intermittent ZDA is detectable.

\subsubsection{\textbf{Full Observation of Velocity}} To proceed with the proof of \eqref{eq:oa1}, we first need to obtain $\ker({\mathcal{O}}_{k})$ of the system \eqref{eq:s2} given in \eqref{eq:om1}. The analysis of the kernel of the observability matrix ${\mathcal{O}}_{k}$ can follow the relation \eqref{eq:pk1} with the setting of $D  = {\mathbf{0}_{\left| \mathbb{M} \right| \times 2\left| \mathbb{V} \right|}}$.
 We note that \eqref{eq:pk1} is equivalently represented by \eqref{eq:fv}, \eqref{eq:fp} and \eqref{eq:po}. The results \eqref{eq:pthb1} and \eqref{eq:aq1} are obtained without considering \eqref{eq:fv1}, \eqref{eq:fp1} and \eqref{eq:po1} which are the only terms involving $D$. Then, results similar to \eqref{eq:pthb1} and \eqref{eq:aq1} can be obtained for the system in \eqref{eq:s2} as
\begin{align}
{{x}_1}\left( {{{\zeta_k}}} \right) \!=\!  \ldots  \!=\! {{x}_{\left| \mathbb{V} \right|}}\left( {{{\zeta_k}}} \right) ~\text{and}~ {{v}_1}\left( {{{\zeta_k}}} \right) \!=\!  \ldots  \!=\! {{v}_{\left| \mathbb{V} \right|}}\left( {{{\zeta_k}}} \right). \label{eq:adde2}
\end{align}
Further, with $D  = {\mathbf{0}_{\left| \mathbb{M} \right| \times 2\left| \mathbb{V} \right|}}$, from \eqref{eq:fv1} with ${\mathbf{v}}\left( {{{\zeta_k}}} \right)$ replaced by ${{v}}\left( {{{\zeta_k}}} \right)$, we have ${C_2}{v}\left( {{\zeta _k}} \right) = {\mathbf{0}_{\left| \mathbb{M} \right|}}$, which combined with
\eqref{eq:adde2} yields ${{x}_1}\left( {{{\zeta_k}}} \right) \!=\!  \ldots  \!=\! {{x}_{\left| \mathbb{V} \right|}}\left( {{{\zeta_k}}} \right)$ and $ {{v}_1}\left( {{{\zeta_k}}} \right) \!=\!  \ldots  \!=\! {{v}_{\left| \mathbb{V} \right|}}\left( {{{\zeta_k}}} \right) = 0$. Thus, $\ker({\mathcal{O}}_{k}) = \left\{ {{{\mathbf{0}}_{2\left| \mathbb V \right|}}}, \left[
    \begin{array}{c;{1pt/1pt}c}
       \!\!\!\mathbf{1}_{\left| \mathbb{V} \right|}^\top \!\!\!&\!\! \mathbf{0}_{\left| \mathbb{V} \right|}^\top \!\!\!
          \end{array}
\right]^\top \right\}$. Since all of the elements in $\ker({\mathcal{O}}_{k})$ are the equilibrium points of the system \eqref{eq:s2}, through the recursive computation of \eqref{eq:ck1} and \eqref{eq:ck2}, we arrive at \eqref{eq:oa1}.

\subsubsection{\textbf{Full Observation of Position}} To obtain $\ker({\mathcal{O}}_{k})$ under full observation of position, we can consider \eqref{eq:fp} with $D = {\mathbf{0}_{\left| \mathbb{M} \right| \times 2\left| \mathbb{V} \right|}}$. From \eqref{eq:fp1}  and \eqref{eq:pthb1} we have ${{x}_1}\left( {{{\zeta_k}}} \right) =  \ldots  = {{x}_{\left| \mathbb{V} \right|}}\left( {{{\zeta_k}}} \right) = 0$. Then, we obtain from \eqref{eq:t2} (replace ${\mathbf{v}_i}\left( {{{\zeta_k}}} \right)$ by ${{v}_i}\left( {{{\zeta_k}}} \right)$) that  $\ker({\mathcal{O}}_{k}) = \left\{ {{{\mathbf{0}}_{2\left| \mathbb V \right|}}} \right\}$, which means that if the monitored agents output full observation of positions, the system \eqref{eq:s2} is observable at $t_{k}$; thus \eqref{eq:oa2} is obtained by the recursive computation of \eqref{eq:ck1} and \eqref{eq:ck2}.

\vspace{-0.1cm}
\subsection{Under Partial Observation}
\vspace{-0.10cm}
The analysis of observability follows the same steps of that under full observation. With $C_1 =  C_2$, from  \eqref{eq:po3} we have ${C_2}\mathcal{L}_r^{d + 1}\mathbf{x}\!\left( {{\zeta _k}} \right) = 0,\forall d \in {\mathbb{N}_0}$. Employing the same steps to derive \eqref{eq:pthb1} under full observation of velocity, we obtain \eqref{eq:pthb1} as well under partial observation. Moreover, substituting \eqref{eq:pthb1} into \eqref{eq:po2} and repeating the same steps,  we arrive at \eqref{eq:aq1}. It is straightforward to verify from the dynamics \eqref{eq:s2} that  ${\mathbf{x}_1}\left( {{{t}}} \right) =  \ldots  = {\mathbf{x}_{\left| \mathbb{V} \right|}}\left( {{{t}}} \right)$ and ${\mathbf{v}_1}\left( {{{t}}} \right) =  \ldots  = {\mathbf{v}_{\left| \mathbb{V} \right|}}\left( {{{t}}} \right)$ for any $t \geq t_{0}$, if and only if \eqref{eq:aq1} and \eqref{eq:pthb1} hold. Finally, considering \eqref{eq:po1} with the setting of $D = {\mathbf{0}_{\left| \mathbb{M} \right| \times 2\left| \mathbb{V} \right|}}$, we have ${C_1}\mathbf{x}\left( {{\zeta _k}} \right) + {C_2}\mathbf{v}\left( {{\zeta _k}} \right) = {\mathbf{0}_{\left| \mathbb{M} \right|}}$, from which we have $\ker({\mathcal{O}}_{k}) = \left\{ {{{\mathbf{0}}_{2\left| \mathbb V \right|}}}, \left[
    \begin{array}{c;{1pt/1pt}c}
       \!\!\!\mathbf{1}_{\left| \mathbb{V} \right|}^\top \!\!\!&\!\! -\mathbf{1}_{\left| \mathbb{V} \right|}^\top \!\!\!
          \end{array}
\right]^\top \right\}, \forall k \in \mathbb{N}_{0}$, and then \eqref{eq:oa1k} is obtained by computation of \eqref{eq:ck1} and \eqref{eq:ck2}.

Under the condition \eqref{eq:nad1}, $\mathbf{z}\left( {{\zeta _k}} \right) \in \mathbf{N}^{k}_{0}$, which in conjunction with \eqref{eq:ss1} implies $\eta_{r} = -1$. Substituting \eqref{eq:pthb1}, \eqref{eq:aq1} and $\eta_{r} = -1$ into \eqref{eq:ss2} yields ${\bar{g}}\left( \zeta^{-}_k \right) = \mathbf{0}_{\left| \mathbb{V} \right|}$, and consequently, ${\breve{g}}\left( \zeta^{-}_k \right) = \mathbf{0}_{2\left| \mathbb{V} \right|}$. This means that there is no ZDA on the system at $\zeta^{-}_k$, which contradicts the assumption that the attack is applied until $\zeta_k$.

\section*{Appendix F: Proof of Theorem \ref{thm:thdb}}
With the definition of $C_j, j = 1, 2$, in \eqref{eq:u1b}, we can rewrite \eqref{eq:pox} as
\begin{subequations}
\begin{align}
{{\dot{\tilde{e}}}_x}\left( t \right) &= {\tilde{e}_v}\left( t \right), \\
{{\dot{\tilde{e}}}_v}\left( t \right) &= -{\tilde e_v}\left( t \right) \!-\! {{\widehat{\mathcal{L}}}_s}{\tilde{e}_x}\left( t \right) \!-\! \left( {{{\widehat{\mathcal{L}}}_s} \!-\! {\mathcal{L}_s}} \right)x\left( t \right),\\
{\widehat{y}}\left( t \right) - y\left( t \right) &= {C_1}{\tilde e_x}\left( t \right) + {C_2}{\tilde e_v}\left( t \right),t \in \left[ {{t_{k + 1}},{t_{k + 2}}} \right)\label{eq:pp}\\
\tilde{e}_{x}\left( {{t_{{{k + 1}}}}} \right) &= {\mathbf{0}_{\left| \mathbb{V} \right|}}, \tilde{e}_{v}\left( {{t_{{{k + 1}}}}} \right) = {\mathbf{0}_{\left| \mathbb{V} \right|}}.
\end{align}\label{eq:op}\end{subequations}
\vspace{-0.50cm}

We define ${\mathcal{C}} \triangleq \emph{\emph{diag}}\left\{ c_{12}, \ldots , c_{\left| \mathbb{D} \right|2} \right\}$, where the diagonal entries are from $C_2$ defined in \eqref{eq:u1b}. According to \eqref{eq:dfv} and $\left| \mathbb{D} \right|\leq \left| \mathbb{M} \right|$ (implied by \eqref{eq:stcl}), the matrix ${\mathcal{C}}$ is invertible. Now, considering \eqref{eq:sep}, we have\begin{align}
{C}_{2}\left( {{{\widehat{\mathcal{L}}}_{s}} - {{\cal L}_{s}}} \right) \!=\! \left[
    \begin{array}{c;{2pt/2pt}c}
        \!\!\!\mathcal{C}\mathfrak{L}_{s} \!\!\!&\!\! {{\mathbf{0}_{\left| \mathbb{D} \right| \!\times\! \left( {\left| \mathbb{M} \right| - \left| \mathbb{D} \right|} \right)}}} \!\!\!\!\\ \hdashline[2pt/2pt]
        \!\!\!{\mathbf{0}_{\left( {\left| \mathbb{M} \right| - \left| \mathbb{D} \right|} \right) \!\times\! \left| \mathbb{D} \right|}}  \!\!\!&\!\! {\mathbf{0}_{\left( {\left| \mathbb{M} \right| - \left| \mathbb{D} \right|} \right) \!\times\! \left( {\left| \mathbb{M} \right| - \left| \mathbb{D} \right|} \right)}}\!\!\!\!
    \end{array}
\right],\label{eq:xxza1c}
\end{align}
which, in conjunction with invertible matrix ${\mathcal{C}}$ and the definitions of ${A_s}$ in \eqref{eq:nm0} and ${{\widehat {A}}_s}$ in \eqref{eq:A_hat}, implies that if ${C}_{2}\left( {{{\widehat{\mathcal{L}}}_{s}} - {{\cal L}_{s}}} \right) {x^{\left( d \right)}}\left( {{t_{k + 1}}} \right) = {\mathbf{0}_{\left| \mathbb{M} \right|}}, \forall d \in {\mathbb{N}_0}$, then
\begin{align}
\left( {{{\widehat {A}}_s} - {A_s}} \right){z^{\left( d \right)}}\left( {{t_{k + 1}}} \right) = {\mathbf{0}_{2\left| \mathbb{V} \right|}}, \forall d \in {\mathbb{N}_0}.\label{eq:xxza1ca}
\end{align}

Under the dynamics \eqref{eq:op} and the relation \eqref{eq:xxza1ca}, the necessary condition \eqref{eq:xxz} of guaranteeing stealthy property of cooperative ZDA is equivalently written as \begin{subequations}
\begin{align}
{C}_{2}\left( {{{\widehat{\mathcal{L}}}_s} - {\mathcal{L}_s}} \right)\mathcal{L}_s^dx\left( t_{k + 1} \right) &= {\mathbf{0}_{\left| \mathbb{M} \right|}},\forall d \in {\mathbb{N}_0}\\
{C}_{2}\left( {{{\widehat{\mathcal{L}}}_s} - {\mathcal{L}_s}} \right)\mathcal{L}_s^dv\left( t_{k + 1} \right) &= {\mathbf{0}_{\left| \mathbb{M} \right|}},\forall d \in {\mathbb{N}_0}.
\end{align}\label{eq:xxza1d}\end{subequations}

\vspace{-0.55cm}
We assume that the topology attack in system \eqref{eq:sas} can ensure that the stealthy property \eqref{eq:rs1} of ZDA holds. Noticing \eqref{eq:xxza1c} and the dynamics \eqref{eq:s2}, the equation \eqref{eq:xxza1d} is equivalent to
${\mathcal{C}}{{\mathfrak{L}}_{\sigma \left({{t_{k + 1}}} \right)}}{\chi ^{\left( m \right)}}(t_{k + 1}) = {\mathbf{0}_{\left| \mathbb{D} \right|}},\forall m \in {\mathbb{N}_0}$, where $\chi(t_{k + 1}) \triangleq \left[\!\!\!
    \begin{array}{c;{1pt/1pt}c;{1pt/1pt}c}
        {{{x}_1}( t_{k + 1})} & {\ldots} & {{x}_{\left| \mathbb{D} \right|}}(t_{k + 1})
    \end{array}
\!\!\!\right]^\top.$ Since ${\mathcal{C}}$ is invertible, we have
\begin{align}
{{\mathfrak{L}}_{\sigma \left({{t_{k + 1}}} \right)}}{\chi ^{\left( m \right)}}(t_{k + 1}) = {\mathbf{0}_{\left| \mathbb{D} \right|}},\forall m \in {\mathbb{N}_0}. \label{eq:tro}
\end{align}

As ${{\mathfrak{L}}_{\sigma \left({{t_{k + 1}}} \right)}}$ is the elementary row transformation of a Laplacian matrix, there exists an elementary row operator $E \in \mathbb{R}^{\left| \mathbb{D} \right| \times \left| \mathbb{D} \right|}$ such that $\widehat{\mathfrak{L}}_{\sigma \left({{t_{k + 1}}} \right)} \triangleq E {{\mathfrak{L}}_{\sigma \left({{t_{k + 1}}} \right)}}$ is a Laplacian matrix. Pre-multiplying both sides of \eqref{eq:tro} by $E$ yields
\begin{align}
\widehat{\mathfrak{L}}_{\sigma \left({{t_{k + 1}}} \right)}{\chi ^{\left( m \right)}}(t_{k + 1}) = {\mathbf{0}_{\left| \mathbb{D} \right|}},\forall m \in {\mathbb{N}_0}. \label{eq:trotras}
\end{align}
It is well-known that the null space of the Laplacian matrix of a connected graph is spanned by the vector with all ones. From \eqref{eq:trotras} we conclude that  $\exists i,j \in \mathbb{D}: x_i^{\left( m \right)}(t_{k + 1}) = x_j^{\left( m \right)}(t_{k + 1}),t_{k + 1} \ge {t_0},\forall m \in {\mathbb{N}_0}$,
which can be rewritten as
\begin{align}
\left( {{\mathrm{e}^\top_i} - {\mathrm{e}^\top_j}} \right){x^{\left( m \right)}}\left( t_{k+1} \right) = 0,\forall m \in {\mathbb{N}_0} \label{eq:trox}
\end{align}
where $\mathrm{e}_{i}$ denotes a vector of length $\left| \mathbb{D} \right|$ with a single nonzero entry with value 1 in its $i$th position.

Due to the dynamics \eqref{eq:s2}, the equation \eqref{eq:trox} leads to
  \begin{subequations}
\begin{align}
\left( {{\mathrm{e}^\top_i} - {\mathrm{e}^\top_j}} \right)\mathcal{L}_{r}^m x\left( t_{k + 1} \right) &= 0,\forall m \in {\mathbb{N}_0}\\
\left( {{\mathrm{e}^\top_i} - {\mathrm{e}^\top_j}} \right)\mathcal{L}_{r}^m v\left( t_{k + 1} \right) &= 0,\forall m \in {\mathbb{N}_0}.
\end{align}\label{eq:fpc}\end{subequations}

\vspace{-0.40cm}
It follows from \eqref{eq:ufo4} that ${\cal L}_r^d = {Q_r}\Lambda _r^dQ_r^ \top$ with $\Lambda _r$ given in \eqref{eq:wp3}, substituting which into
\eqref{eq:fpc} yields that for $\forall m \in \mathbb{N}$,
  \begin{subequations}
\begin{align}
\!\!\!\sum\limits_{l = 2}^{\left| \mathbb{V} \right|}\! {\lambda _l^m\!({{\cal L}_r})\!\left( {{{\left[ {{Q_r}} \right]}_{i,l}} \!\!-\! {{\left[ {{Q_r}} \right]}_{j,l}}} \right)\!\!\left[ {{Q^\top_r}} \right]_{l,:}{x}\!\left( {{t_{k + 1}}} \right)}  &\!=\! 0, \label{eq:fpthz1}\\
\!\!\!\sum\limits_{l = 2}^{\left| \mathbb{V} \right|}\! {\lambda _l^m\!({{\cal L}_r})\!\left( {{{\left[ {{Q_r}} \right]}_{i,l}} \!\!-\! {{\left[ {{Q_r}} \right]}_{j,l}}} \right)\!\!\left[ {{Q^\top_r}} \right]_{l,:}{v}\!\left( {{t_{k + 1}}} \right)} &\! = \!0.\label{eq:fpthz2}
\end{align}\end{subequations}
Then, with the definitions
\begin{align}
\!\!\!{\mathcal{D}_{ij}} &\!\triangleq\! \emph{\emph{diag}}\!\left\{\! {{{\left[ {{Q_r}} \right]}_{i,2}} - {{\left[ {{Q_r}} \right]}_{j,2}}, \ldots ,{{\left[ {{Q_r}} \right]}_{i,\left| \mathbb{V} \right|}} - {{\left[ {{Q_r}} \right]}_{j,\left| \mathbb{V} \right|}}} \!\right\}\!,\label{eq:impod0}\\
\!\!\!f &\!\triangleq\! \left[ {{Q_r}} \right]_{{2:\left| \mathbb{V} \right|}, :}^ \top x\left( {{t_{k + 1}}} \right), \label{eq:impod}
\end{align}
following the same derivations from \eqref{eq:xxc} to \eqref{eq:ptha1}, we arrive at
\begin{align}
{\widetilde{\mathcal{H}}_r}\mathcal{D}_{ij}f = {\mathbf{0}_{\left| \mathbb{V} \right|-1}}, \forall i \in \mathbb{M},\label{eq:tro3}
\end{align}
where $\widetilde{\mathcal{H}}_r$ is given in \eqref{eq:pth27}. Using the same analysis to derive \eqref{eq:ptha2}, we conclude that under the condition \eqref{eq:kkz0}, the solution of \eqref{eq:tro3} is $\mathcal{D}_{ij}f = {\mathbf{0}_{\left| \mathbb{V} \right|-1}}$. Since ${\mathcal{D}_{ij}}$ given by \eqref{eq:impod0} is full-rank under the condition \eqref{eq:kkz1x}, we have $f = {\mathbf{0}_{\left| \mathbb{V} \right|-1}}$. Then, noticing \eqref{eq:impod}, by Lemma~\ref{thm:ndd} we arrive at
\begin{align}
{{x}_1}\left( {{{t_{k+1}}}} \right) =  \ldots  = {{x}_{\left| \mathbb{V} \right|}}\left( {{{t_{k+1}}}} \right). \label{eq:fpthz3}
\end{align}
Repeating the same procedure of deriving \eqref{eq:fpthz3} from \eqref{eq:fpthz1}, we conclude  ${{v}_1}\left( {{{t_{k+1}}}} \right) =  \ldots  = {{v}_{\left| \mathbb{V} \right|}}\left( {{{t_{k+1}}}} \right)$ from \eqref{eq:fpthz2}, which means that the second-order consensus is achieved at $t_{k + 1}$, i.e.,  ${x_i}\left( {{t_{k + 1}}} \right) = {x_j}\left( {{t_{k + 1}}} \right)$ and ${v_i}\left( {{t_{k + 1}}} \right) = {v_j}\left( {{t_{k + 1}}} \right)$, $\forall i \ne j \in \mathbb{V}$. It is straightforward to verify from the dynamics \eqref{eq:foon} that the second-order consensus is achieved at some time $t < \infty$ if and only if the individual initial conditions are identical, i.e., ${x_i}\left( {{t_{0}}} \right) = {x_j}\left( {{t_{0}}} \right)$ and ${v_i}\left( {{t_{0}}} \right) = {v_j}\left( {{t_{0}}} \right)$.  Hence, the cooperative ZDA is undetectable only in the case of identical initial condition that corresponds to the steady state.

\section*{Appendix G: Proof of Theorem \ref{thm:dfd}}
We define ${\mathbf{e}_x}\left( t \right) \triangleq q\left( t \right) - \breve{x}\left( t \right)$ and ${\mathbf{e}_v}\left( t \right) \triangleq w\left( t \right) - \breve{v}\left( t \right)$. The dynamics of tracking errors in the presence of the attack  obtained from \eqref{eq:fl} and \eqref{eq:oofn} are given as:
\begin{subequations}
\begin{align}
\!\!\!{{\dot{\mathbf{e}}}_{x_{i}}}\!\!\left( t \right) &\!=\! {\mathbf{e}_{v_{i}}}\!\!\left( t \right), \\
\!\!\!{{\dot{\mathbf{e}}}_{v_{i}}}\!\!\left( t \right) &\!=\! - {{\bf{e}}_{v_{i}}}\!\!\left( t \right) + \sum\limits_{i \in \mathbb{V}} \!{a_{ij}^{\sigma \left( t \right)}}\!\!\left( {{{{\mathbf{e}}}_{x_{j}}}\!\!\left( t \right) - {{{\mathbf{e}}}_{x_{i}}}\!\!\left( t \right)} \right)  \nonumber\\
 &\hspace{0.3cm}- \!\left\{ \begin{array}{l}
\hspace{-0.2cm}{\breve{g}}_i\!\left( t \right)\!, i \!\in\! \mathbb{K}\\
\hspace{-0.2cm}0, \hspace{0.10cm} i \!\in\! {\mathbb{V}} \backslash \mathbb{K}
\end{array} \right. \!\!\!\!-\! \left\{ \begin{array}{l}
\hspace{-0.2cm}r_{i}(t), \hspace{0.92cm}c_{i1} \!\neq\! 0, \!i \!\in\! \mathbb{M}\\
\hspace{-0.2cm}\int_{t_{0}}^{t}\!{r_{i}(b)}\mathrm{d}b,\hspace{0.1cm} c_{i1} \!=\! 0, \! i \!\in\!  \mathbb{M}\\
\hspace{-0.2cm}0, \hspace{1.4cm} i \!\in\! {\mathbb{V}} \backslash \mathbb{M}\\
\end{array} \right. \\
\!\!\!r_{i}\!\left( t \right) &\!=\! {c_{i1}}{{\bf{e}}_{{x_i}}}\!\!\left( t \right)  + {c_{i2}}{{\bf{e}}_{{v_i}}}\!\!\left( t \right) - {d_i}{\breve{g}_i}\!\left( t \right),i \in \mathbb{M}.
\end{align}\label{eq:fab1}\end{subequations}

The attack is not detected by the observer \eqref{eq:fl} means that $r_{i}\left( t \right) = 0, i \in \mathbb{M}$, for any $t \geq t_{0}$. Substituting it into the above equation results in
\begin{subequations}
\begin{align}
{{\dot{\mathbf{e}}}_{x_{i}}}\!\!\left( t \right) &= {{{\mathbf{e}}}_{v_{i}}}\!\!\left( t \right) \nonumber\\
{{\dot{\mathbf{e}}}_{v_{i}}}\!\!\left( t \right) &=-{{{\mathbf{e}}}_{v_{i}}}\!\!\left( t \right) + \sum\limits_{i \in \mathbb{V}} \!{a_{ij}^{\sigma \left( t \right)}}\!\!\left( {{{{\mathbf{e}}}_{x_{j}}}\!\!\left( t \right) - {{{\mathbf{e}}}_{x_{i}}}\!\!\left( t \right)} \right) - \left\{ \begin{array}{l}
\hspace{-0.2cm}{\breve{g}}_i\!\left( t \right)\!, i \!\in\! \mathbb{K}\\
\hspace{-0.2cm}0, \hspace{0.10cm} i \!\in\! {\mathbb{V}} \backslash \mathbb{K}
\end{array} \right. \nonumber\\
r_{i}\!\left( t \right) &= {c_{i1}}{{\bf{e}}_{{x_i}}}\!\!\left( t \right) + {c_{i2}}{{\bf{e}}_{{v_i}}}\!\!\left( t \right) - {d_i}{\breve{g}_i}\!\left( t \right),i \in \mathbb{M} \nonumber
\end{align}\end{subequations}
which has the same form of dynamics as that of \eqref{eq:oofn}. Therefore, the analysis of ZDA variations in the observer \eqref{eq:fl} follows the same analysis of the system \eqref{eq:oofn}. Moreover, the required condition \eqref{eq:dfv} implies that the monitored agents output full observations of velocity or partial observations: either \eqref{eq:oa1} or \eqref{eq:oa1k} implies \eqref{eq:ccd}. Hence, the topology attacker cannot infer the real-time full states of the non-monitored agents, and the topology attacker has to consider the scope of the target connections  implied by \eqref{eq:stcl}. Therefore, the proof of the first statement follows from Theorems~\ref{thm:thd} and \ref{thm:thdb}.

In the absence of attacks, the system matrix of system \eqref{eq:fab1} is $\widehat{\mathcal{A}}_{\sigma(t)}$ defined in \eqref{eq:dm}.   Since the condition \eqref{eq:kkz0} implies that all of the switching topologies provided to Algorithm~1 are connected graphs and condition \eqref{eq:dfv} implies \eqref{eq:cog}, the matrix $\widehat{\mathcal{A}}_{\sigma(t)}$ is Hurwitz by Lemma~\ref{thm:my0bd}. Thus, there exists a $P > 0$ such that both \eqref{eq:sdsbk1} and \eqref{eq:sdsbk10} hold. Hence, the proof of the second statement follows from Proposition~\ref{thm:zzaxx} and Lemma \ref{thm:my0bd}.
\bibliographystyle{IEEEtran}
\bibliography{ref}
\end{document}